\documentclass[12pt, axodraw]{article}
\usepackage{graphicx, axodraw,amssymb, amsmath}
\newcommand{\mathsym}[1]{{}}

\input epsf.tex

\let\badcite=\cite
\def\cite{~\badcite}

\def\slashchar#1{\setbox0=\hbox{$#1$}           
   \dimen0=\wd0                                 
   \setbox1=\hbox{/} \dimen1=\wd1               
   \ifdim\dimen0>\dimen1                        
      \rlap{\hbox to \dimen0{\hfil/\hfil}}      
      #1                                        
   \else                                        
      \rlap{\hbox to \dimen1{\hfil$#1$\hfil}}   
      /                                         
   \fi}
    \def\slashword#1{\setbox0=\hbox{$#1$}        
  \dimen0=\wd0                                   
   \setbox1=\hbox{/} \dimen1=\wd1                
   \ifdim\dimen0>\dimen1                         
      \rlap{\hbox to \dimen0{\hfil\bf---\hfil}} %
      #1                                         %
   \else                                         
      \rlap{\hbox to \dimen1{\hfil$#1$\hfil}}    
      /                                          
    \fi}                                         %

\catcode`@=11
\newdimen\vbigd@men                             

\def\vbig#1#2{{\vbigd@men=#2\divide\vbigd@men by 2%
   \hbox{$\left#1\vbox to \vbigd@men{}\right.\n@space$}}}
\catcode`@=12

\catcode`@=11
\def\citenum#1{\csname b@#1\endcsname}
\catcode`@=12

\begin{document}
\begin{titlepage}

\begin{flushright}
{SCUPHY-TH-08001}\\
\end{flushright}

\bigskip\bigskip

\begin{center}{\Large\bf\boldmath
Extracting MSSM Masses From Heavy Higgs Decays to Four Leptons at the LHC }
\end{center}
\bigskip
\centerline{\bf P. Huang, N. Kersting\footnote{Email: nkersting@scu.edu.cn}, H.H. Yang }
\centerline{{\it Physics Department, Sichuan University, P.R. China 610065}}

\bigskip

\begin{abstract}
It is well known that finding and measuring the masses of particles in the Minimal Supersymmetric Standard Model (MSSM) at the Large Hadron Collider (LHC) may be possible using invariant mass distributions in exclusive channels containing $n_j$ jets and  $n_\textit{l}$ leptons. We perform this analysis for the $(n_j, n_\textit{l}) = (0,4)$ decay of heavy Higgs bosons to neutralinos, 
$pp \to H/A \to \widetilde{\chi}_{i}^0 \widetilde{\chi}_{j}^0$ ($i,j =2,3,4$), which then decay to four leptons and two lightest neutralinos $\widetilde{\chi}_{1}^0$  via on-shell sleptons. When $i=j$ and the sleptons are degenerate, our Monte Carlo study shows that the LHC will be able to measure the Higgs and relevant neutralino and slepton masses to roughly 
 30\%; however, if one of these is  already known within 5\%, the other three may be found to equal or better accuracy. This would provide the first accurate measurement of the $H/A$ mass via invariant mass distribution techniques.
\end{abstract}

\newpage
\pagestyle{empty}

\end{titlepage}


\section{Introduction}

Supersymmetry (SUSY) provides one of the most attractive candidates for physics beyond the Standard Model (SM) anticipated at the Large Hadron Collider (LHC). The Minimal SUSY SM (MSSM), in particular, offers a rich spectrum of novel super-particles (sparticles) which,
 if relevant to LHC phenomenology, will have masses in the range of hundreds of GeV.
Yet measuring this spectrum will be far from trivial since the center-of-mass (CM) energy of the LHC is available to us only by partonic interactions and therefore not fixed from event to event: we will not be able to tune the CM energy to precisely scan through mass resonances. The general consensus in the literature is to first conduct inclusive measurements\cite{inc1,inc2,inc3,inc4,inc5} which require a suitable number of $n_j$ high energy jets plus $n_\textit{l}$ isolated leptons plus missing energy (as carried out of the detector system by the lightest SUSY particle (LSP) in R-partity-conserving models), permitting a gross measurement of the sparticles' mass scale, and subsequently specialize to analysis of exclusive channels.
 Performing the latter in as model-independent a way as possible is a challenge in light of the tremendous parametric freedom in the MSSM, where sparticle decay topologies vary widely across this parameter space.

  Once a specific decay chain can be identified we can construct relativistically-invariant combinations of $n_j$ and $n_\textit{l}$ momenta and analyze their distributions over a large number of events.
Since endpoints of these distributions are typically well-defined analytic functions of sparticle masses for this given decay topology, a sufficiently large and pure sample of sparticle events can provide a clean endpoint which  constrains some set of sparticle masses. Examples of situations where this technique yields promising results include where we can identify
$\widetilde{\chi}_{2}^0  \to  {l}^\pm {l}^\mp \widetilde{\chi}_{1}^0 $ (one endpoint for two unknown sparticle masses),
$ \widetilde{l_L}^\pm \to  \widetilde{\chi}_{2}^0 {l}^\pm \to
\widetilde{l'_R}^\pm {l'}^\mp {l}^\pm   \to
{l}^\pm  {l'}^\pm {l'}^\mp    \widetilde{\chi}_{1}^0$ (one endpoint for three masses),
$pp \to \tilde{q}_L \tilde{q}_L $  with $\tilde{q}_L \to q \widetilde{\chi}_{2}^0
\to \tilde{l}^\pm {l}^\mp q \to {l}^\pm {l}^\mp q \widetilde{\chi}_{1}^0$ or
$ \tilde{q}_L \to q \widetilde{\chi}_{2}^0  \to
 h q \widetilde{\chi}_{1}^0 $ (six endpoints for four masses ), and $\tilde{g} \to \tilde{q} q \to \widetilde{\chi}_{2}^0 q q \to
 \tilde{l}^\pm {l}^\mp q q \to l^\pm {l}^\mp q q \widetilde{\chi}_{1}^0$ (seven endpoints for five masses), all of which are discussed in more detail in Refs.\cite{invmass1},\cite{invmass2}, and\cite{invmass3}.
However, in these examples either the number of endpoints is less than the number of unknown masses, so we can only constrain the MSSM to some \textit{surface} in mass parameter space, and/or we have to make extra model assumptions which ensure the assumed decay topology occurs with a sufficient rate.

In this paper we investigate whether this technique performs better with a larger number $n_\textit{l}$ of final state leptons. As signals consisting of isolated leptons provide cleaner signals than those from jets at a hadron collider, invariant masses constructed from the former may be subject to less error.
 While up to now researchers have considered
$n_\textit{l}=3$ which only gives three independent invariant combinations, higher values of $n_\textit{l}$  rapidly give more:   $n_\textit{l}=4$ already gives seven combinations\footnote{For an endstate with  $n_j$ jets and $n_\textit{l}$ leptons ($N \equiv n_j + n_\textit{l}$), the number of independent invariant combinations is equal to the number of pairwise contractions, `N choose 2', plus the number of contractions with  $\epsilon_{\mu\nu\rho\sigma}$, `N choose 4'. $N=4$, $5$, and $6$ gives $7$, $15$, and $30$ combinations, respectively.}.
 For virtually any conceivable MSSM decay chain this would already suffice to overconstrain the unknown masses if the endpoints of these distributions were precisely known. However, with this rise of the number of constraints comes
  the lower statistics on each constraint since  cross-sections naturally fall with $n_\textit{l}$. But backgrounds (both SUSY and SM) will correspondingly become smaller, hence confidence in the assumed decay channel will increase. Moreover, there is information in the peaks of the distributions as well -- by definition these have higher statistics and therefore lower error, though we would have to investigate the  sensitivity of these (or any other `shape' variable) to cuts.
  A complex interplay of all these factors therefore determines the success of this programme, which ultimately depends on the specific endstate considered.

In this work we concentrate on the ($n_j=0$,  $n_\textit{l}=4$) endstate which
may result from the decays of heavy Higgs bosons $H^0$ and $A^0$ (hereafter collectively referred to as `Higgs') to neutralinos,
\begin{equation} \label{hdecay}
pp \to H/A \to \widetilde{\chi}_{i}^0 \widetilde{\chi}_{j}^0 \to
    \widetilde{l_1}^\pm {l_1}^\mp \widetilde{l_2}^\pm {l_2}^\mp
    \to {l_1}^\mp {l'_1}^\pm {l_2}^\mp {l'_2}^\pm \widetilde{\chi}_{1}^0 \widetilde{\chi}_{1}^0
\end{equation}
proceeding via on-shell sleptons of electron or muon ($l_{1,2} \subset \{e,\mu\}$) flavor. As
shown in Ref.\cite{ha4l} this signal has favorable rates ($\geq 100$ events for
$100\, \hbox{fb}^{-1}$ integrated luminosity at the LHC) over much of the $(\mu,M_2)$-plane\footnote{Gauge unification here fixes $M_1 = 5/3 \tan^2 \theta_W M_2$.} when the Higgs mass is in a favorable range
($350\, \hbox{GeV} < M_A < 700\, \hbox{GeV}$) and first and second generation slepton masses are sufficiently light ($<200\, \hbox{GeV}$).
Here the correlation between lepton invariant mass pairs, $i.e.$ a wedgebox plot\cite{EWwedgebox,cascade}, tells us something about the mass differences
 $m_{\widetilde{l_1}} - m_{\widetilde{l_2}}$ and $m_{{\widetilde\chi}^0_i}- m_{{\widetilde\chi}^0_j}$: they are zero for a symmetric boxlike wedgebox plot.
 In this ($i=j$) case  the only significant background after a suitable jet cut consists of charginos\footnote{One might worry about neutralino channels $pp \to \widetilde{\chi}^0_i \widetilde{\chi}^0_j$, however,  as shown in Ref.\cite{EWwedgebox}, these
require a coupling to the Z-boson which, in the notation of \cite{mixing}, is proportional to $(N_{i3}N_{j3}^* - N_{i4}N_{j4}^*)$; here the crucial minus sign arises from the different hypercharges of the two MSSM Higgs doublets. For $i=j$ this coupling is highly suppressed;  $i \ne j$ processes can be significant, but these do not give a boxlike wedgebox plot.}, $i.e.$ $pp \to  \widetilde{\chi}_{i}^\pm  \widetilde{\chi}_{j}^\mp$, which may subsequently decay as 
\begin{eqnarray*}
 \widetilde{\chi}_{i}^\pm & \to & W^\pm (\to l^\pm ~ \nu) ~  \widetilde{\chi}_{k}^0 
           (\to {l'}^\pm \tilde{l'}^\mp(\to {l'}^\mp \widetilde{\chi}_{1}^0))  \\
  \widetilde{\chi}_{j}^\mp & \to & {l''}^\mp ~ \tilde{\nu}'' (\to \nu'' \widetilde{\chi}_{1}^0 )  
\end{eqnarray*}
Among such four-lepton endstates, the ratio of
 flavor-balanced ($e^+ e^- e^+ e^- + \mu^+ \mu^- \mu^+ \mu^- + e^+ e^- \mu^+ \mu^-$ ) to flavor-unbalanced ($e^+ e^- e^\pm \mu^\mp + \mu^+ \mu^- \mu^\pm e^\mp$) events, which we define as
$\mathcal{R}_\pm$,  is close to unity since
 there is no correlation between the flavors of ${l}$ and ${ l''}$.
On the other hand, a pure Higgs signal would have no flavor-unbalanced events and hence $\mathcal{R}_\pm \to \infty$. A boxlike wedgebox plot, therefore, not only 
contains much information in leptonic invariant mass distributions (potentially seven upper and seven lower endpoints, if these are indeed independent, plus constraints from seven peaks) which can be used to find the six unknown masses with high precision, but also provides a model-independent  confidence measure ($ \mathcal{R}_\pm$).

The rest of this paper is organized as follows: in Section \ref{sec:th} we present our derivation of analytical expressions for the endpoints of the seven invariant combinations of lepton momenta (since exact formulae are rather lengthy they are collected in the Appendix); in Section \ref{sec:mc} we test this method with Monte Carlo (MC) simulation of LHC data generated at three different MSSM parameter points with backgrounds and detector effects included. Section \ref{sec:sumdisc}
summarizes and discusses these results.

\section{Endpoint Theory}
\label{sec:th}

\begin{figure}[]
{
\unitlength=1.3 pt
\SetScale{1.25}
\SetWidth{0.5}      
\scriptsize    
\begin{picture}(200,100)(0,-100)
\ArrowLine(180,-40)(250,-40)
\ArrowLine(180,-40)(110,-40)
\BCirc(180,-40){8}
\Text(173.5,-39)[c]{\tiny \bf $H/A$}
\Text(200,-30)[c]{ \bf $\widetilde{\chi}_{i}^0$}
\Text(150,-30)[c]{ \bf $\widetilde{\chi}_{j}^0$}
\ArrowLine(240,-40)(205,-80)
\Text(220,-43)[c]{ \bf $\theta_i$}
\CArc(240,-40)(18,180,228)
\ArrowLine(120,-40)(155,-80)
\Text(125,-43)[c]{ \bf $\theta_j$}
\CArc(120,-40)(18,312,360)
\Text(210,-75)[c]{ \bf $l_1$}
\Text(137,-75)[c]{ \bf $l_2$}
\ArrowLine(240,-40)(275,0)
\ArrowLine(120,-40)(85,0)
\ArrowArcn(109,-40)(5,130,-90)
\Text(100,-40)[c]{ \bf $\phi_j$}
\ArrowArc(251,-40)(5,50,270)
\Text(245,-40)[c]{ \bf $\phi_i$}
\ArrowLine(93.2,-9.8)(130,10)
\ArrowLine(93.2,-9.8)(56.4,-30.2)
\ArrowLine(266.8,-9.8)(230,10)
\ArrowLine(266.8,-9.8)(303.6,-30.2)
\Text(88.2,-15)[c]{ \bf $\theta_2$}
\CArc(93.2,-9.8)(11,208,310)
\Text(258.8,-15)[c]{ \bf $\theta_1$}
\CArc(266.8,-9.8)(11,225,330)
\Text(238,-22)[c]{ \bf $\widetilde{l_1}$}
\Text(108,-22)[c]{ \bf $\widetilde{l_2}$}
\ArrowArc(84,1)(5,180,20)
\Text(80,4)[c]{ \bf $\phi_2$}
\ArrowArcn(274,1)(5,0,170)
\Text(264,4)[c]{ \bf $\phi_1$}
\Text(120,0)[c]{ \bf $\widetilde{\chi}_{1}^0$}
\Text(225,0)[c]{ \bf $\widetilde{\chi}_{1}^0$}
\Text(290,-35)[c]{ \bf $l'_1$}
\Text(55,-35)[c]{ \bf $l'_2$}
\end{picture} \
}
\caption{\it \small Kinematic picture of the Higgs decay showing angular variables defined in the rest frames of the decaying particle at the corresponding vertex.}
 \label{decayfig}
\end{figure}
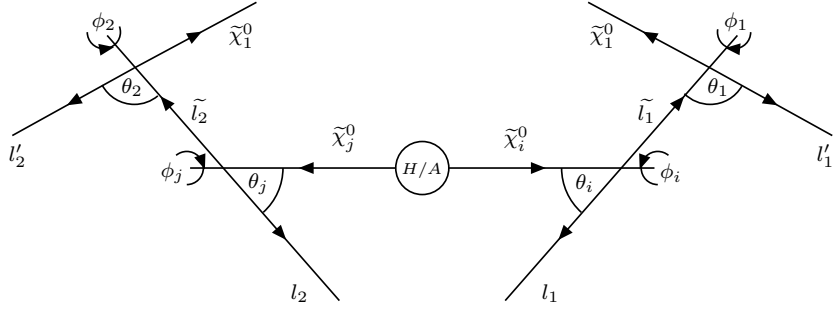

From the four-lepton endstate of (\ref{hdecay}) depicted in Fig.~\ref{decayfig}, we can form six independent relativistically invariant bi-contractions (we take lepton masses to be zero in the following),
\begin{equation}\label{inv}
    (p_{1})^\mu (p_{1'})_\mu~,  ~~~
  (p_{1})^\mu (p_{2})_\mu~, ~~~
  (p_{1})^\mu (p_{2'})_\mu~, ~~~
   (p_{1'})^\mu (p_{2})_\mu~, ~~~
     (p_{1'})^\mu (p_{2'})_\mu~,~~~
  (p_{2})^\mu (p_{2'})_\mu
\end{equation}
in addition to the totally antisymmetric invariant
\begin{equation}\label{asymm4}
   a_4 \equiv  p_{1}^\mu p_{1'}^\nu p_{2}^\rho p_{2'}^\sigma \epsilon_{\mu \nu \rho \sigma}
\end{equation}
However, when we construct a distribution of a function of these invariants we must
take care that this function is totally symmetric under interchanges of
labels $1 \leftrightarrow 1'$ and $2 \leftrightarrow 2'$ because of ambiguity in lepton identification. The usual dilepton invariant masses
\begin{equation}\label{m2l}
    M_{2l} ~\equiv~ 2\,(p_{1})^\mu (p_{1'})_\mu ~~~, ~~~~~~~~
    M_{2l'} ~\equiv~ 2\,(p_{2})^\mu (p_{2'})_\mu
\end{equation}
obey this rule.

But because we want to include same-flavor endstates  ($e^+ e^- e^+ e^- $ and
 $\mu^+ \mu^- \mu^+ \mu^- $)
  in addition to those of opposite flavor ($e^+ e^- \mu^+ \mu^- $), we choose a list of positive definite\footnote{Symmetric {\it linear} combinations of (\ref{inv}) can only be proportional to the four-lepton invariant mass
$M_{4l}$. To get other independent combinations we must, in the simplest scheme, choose averages of squares of  (\ref{inv}). In doing so all lower endpoints become zero.} functions of the invariants in (\ref{inv}) which are symmetric under all label interchanges:
\begin{eqnarray}\label{avinv}
  M_{4l}^2 &\equiv& (p_1 + p_{1'}+ p_2 + p_{2'})^2 \\ \nonumber
  \overline{M}_{{2l2l}}^4 &\equiv& \{(p_1 + p_{1'}- p_2 - p_{2'})^4 +
  (p_1 + p_{2'}- p_2 - p_{1'})^4 + (p_1 + p_{2}- p_{1'} - p_{2'})^4\} /3
     \\  \nonumber
  \overline{M}_{{l3l}}^4 &\equiv&  \{
  (p_1 + p_{1'}+ p_2 - p_{2'})^4 +
  (p_1 + p_{1'}+ p_{2'} - p_2)^4   \\  \nonumber
  &&
  ~~~~~~~~~~~ + (p_1 + p_{2}+ p_{2'} - p_{1'})^4 +
  (p_{2}+ p_{2'}+ p_{1'} - p_{1})^4
 \} /4\\  \nonumber
  \overline{M}_{{l2l}}^4 &\equiv&  \{
  (p_1 + p_{1'}- p_2)^4 +
  (p_1 + p_{1'}- p_{2'} )^4
  + (p_1 + p_{2}- p_{2'})^4 +
  (p_{2}+ p_{2'}- p_{1'} )^4  \\  \nonumber
  && ~~~+
  (p_1 - p_{1'}+ p_2)^4 +
  (p_1 - p_{1'}+ p_{2'} )^4
  + (p_1 - p_{2}+ p_{2'})^4 +
  (p_{2}- p_{2'}+ p_{1'} )^4      \\  \nonumber
 && ~~~+
  (  p_{1'}+ p_2-p_1)^4 +
  (  p_{1'}+ p_{2'}-p_1 )^4
  + (  p_{2}+ p_{2'}-p_1)^4 +
  ( p_{2'}+ p_{1'} -p_{2})^4  \}/12
    \\  \nonumber
  \overline{M}_{{3l}}^4 &\equiv&  \{
  (p_1 + p_{1'}+ p_2)^4 +
  (p_1 + p_{1'}+ p_{2'} )^4
  + (p_1 + p_{2}+ p_{2'})^4 +
  (p_{2}+ p_{2'}+ p_{1'} )^4
 \}/4 \\  \nonumber
  \overline{M}_{{ll}}^4 &\equiv&
   \{
  (p_1 + p_{1'})^4 +
  (p_1 + p_{2'})^4 + (p_1 + p_{2})^4
  + ( p_2 + p_{2'})^4 +
  ( p_2 + p_{1'})^4 + ( p_{1'} + p_{2'})^4 \} /6
  \\  \nonumber
\end{eqnarray}

 With this set of invariants there will be no combinatoric background from lepton misidentification; on the other hand, since the minimum values of (\ref{avinv}) are in fact zero,
we only have upper endpoints (hereafter simply 'endpoints') to measure; the former effect should be more of an advantage than the latter a disadvantage.
Since the seven independent endpoints of  (\ref{asymm4}) and
 (\ref{avinv}) are well-defined functions of the six masses  $m_A$, $ m_{\widetilde{l_1}}$, $m_{\widetilde{l_2}}$,
 $m_{\widetilde{\chi}_{i}^0}$, $m_{\widetilde{\chi}_{j}^0}$, and $m_{\widetilde{\chi}_{1}^0}$, endpoint measurements would plausibly lead to a system of equations which overconstrains these masses.

To derive analytical formulae for the endpoints, we start by defining the kinematic degrees of freedom of the decay chain (\ref{hdecay}) in Fig.~\ref{decayfig}. Since all decays are two-body, daughter particles are produced back-to-back in the rest frame of the decaying particle with spherical angles $(\theta, \phi)$, distributed evenly in the ranges $0 ~\leq~  \phi ~\leq~ 2 \pi$  and $-1 ~\leq~  \cos \theta ~\leq~ 1 $ (we ignore spin effects; this will be justified by comparison with Monte Carlo results below).
The momenta and directions of all four leptons and the LSPs are then completely determined by the set of 8 angles $(\theta_i, \phi_i)$, $(\theta_j, \phi_j)$, $(\theta_1, \phi_1)$, and $(\theta_2, \phi_2)$ which are defined in the rest frames of the $\widetilde{\chi}_{i}^0$, $\widetilde{\chi}_{j}^0$,  $\widetilde{l_1}$, and
$\widetilde{l_2}$, respectively (for exact procedure, please see Appendix).

 Since (\ref{avinv}) define a set of invariant \emph{masses}, they will in general be maximized when some particular sum of lepton energies is maximized while a corresponding sum of 3-momentum is minimized; this type of situation is realized when each polar angle $\theta_i,~\theta_j,~\theta_1,\textrm{and} ~\theta_2$ equals $0$ or $\pi$, values of the various $\phi$ angles then being irrelevant. Each expression in (\ref{avinv}) therefore has potentially $2^4=16$ possible endpoints (though some of these are the same by symmetry), the maximum (true endpoint) of which depends on
the precise values of ratios among the masses $m_A , ~ m_{\widetilde{l_1}}, ~m_{\widetilde{l_2}},
 ~m_{\widetilde{\chi}_{i}^0}, ~m_{\widetilde{\chi}_{j}^0}, {\rm and} ~m_{\widetilde{\chi}_{1}^0}$.

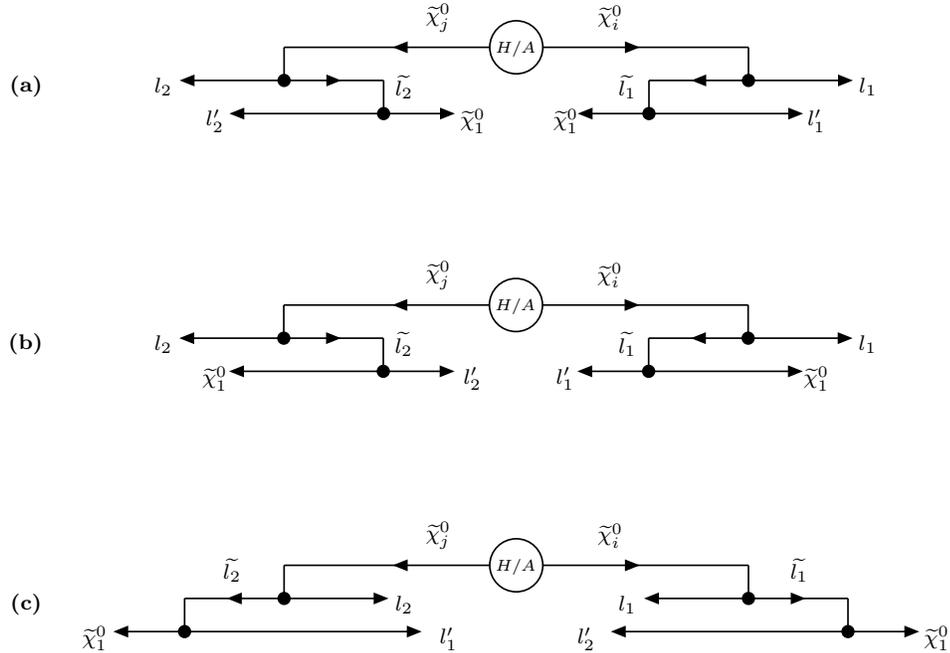
\begin{figure}[]
{
\unitlength=1.3 pt
\SetScale{1.25}
\SetWidth{0.5}      
\scriptsize    
\begin{picture}(200,130)(0,-150)
\ArrowLine(180,-40)(250,-40)
\ArrowLine(180,-40)(110,-40)
\BCirc(180,-40){8}
\Text(173.5,-39)[c]{\tiny \bf $H/A$}
\Text(200,-30)[c]{ \bf $\widetilde{\chi}_{i}^0$}
\Text(150,-30)[c]{ \bf $\widetilde{\chi}_{j}^0$}
\LongArrow(250,-50)(280,-50)
\ArrowLine(250,-50)(220,-50)
\Vertex(250,-50){2}
\LongArrow(110,-50)(80,-50)
\ArrowLine(110,-50)(140,-50)
\Vertex(110,-50){2}
\LongArrow(220,-60)(265,-60)
\LongArrow(220,-60)(200,-60)
\Vertex(220,-60){2}
\LongArrow(140,-60)(160,-60)
\LongArrow(140,-60)(95,-60)
\Vertex(140,-60){2}
\Text(160,-60)[c]{ \bf $\widetilde{\chi}_{1}^0$}
\Text(187,-60)[c]{ \bf $\widetilde{\chi}_{1}^0$}
\Text(205,-50)[c]{ \bf $\widetilde{l_1}$}
\Text(140,-50)[c]{ \bf $\widetilde{l_2}$}
\Text(275,-50)[c]{ \bf $l_1$}
\Text(70,-50)[c]{ \bf $l_2$}
\Text(260,-60)[c]{ \bf $l'_1$}
\Text(85,-60)[c]{ \bf $l'_2$}
\Line(250,-40)(250,-50)
\Line(110,-40)(110,-50)
\Line(220,-50)(220,-60)
\Line(140,-50)(140,-60)
\Text(30,-50)[c]{ \bf (a)}
\end{picture} \
}
{
\unitlength=1.3 pt
\SetScale{1.25}
\SetWidth{0.5}      
\scriptsize    
\begin{picture}(200,60)(211,-75)
\ArrowLine(180,-40)(250,-40)
\ArrowLine(180,-40)(110,-40)
\BCirc(180,-40){8}
\Text(173.5,-39)[c]{\tiny \bf $H/A$}
\Text(200,-30)[c]{ \bf $\widetilde{\chi}_{i}^0$}
\Text(150,-30)[c]{ \bf $\widetilde{\chi}_{j}^0$}
\LongArrow(250,-50)(280,-50)
\ArrowLine(250,-50)(220,-50)
\Vertex(250,-50){2}
\LongArrow(110,-50)(80,-50)
\ArrowLine(110,-50)(140,-50)
\Vertex(110,-50){2}
\LongArrow(220,-60)(265,-60)
\LongArrow(220,-60)(200,-60)
\Vertex(220,-60){2}
\LongArrow(140,-60)(160,-60)
\LongArrow(140,-60)(95,-60)
\Vertex(140,-60){2}
\Text(260,-60)[c]{ \bf $\widetilde{\chi}_{1}^0$}
\Text(85,-60)[c]{ \bf $\widetilde{\chi}_{1}^0$}
\Text(205,-50)[c]{ \bf $\widetilde{l_1}$}
\Text(140,-50)[c]{ \bf $\widetilde{l_2}$}
\Text(275,-50)[c]{ \bf $l_1$}
\Text(70,-50)[c]{ \bf $l_2$}
\Text(160,-60)[c]{ \bf $l'_2$}
\Text(187,-60)[c]{ \bf $l'_1$}
\Line(250,-40)(250,-50)
\Line(110,-40)(110,-50)
\Line(220,-50)(220,-60)
\Line(140,-50)(140,-60)
\Text(30,-50)[c]{ \bf (b)}
\end{picture} \
}
{
\unitlength=1.3 pt
\SetScale{1.25}
\SetWidth{0.5}      
\scriptsize    
\begin{picture}(200,60)(0,-60)
\ArrowLine(180,-40)(250,-40)
\ArrowLine(180,-40)(110,-40)
\BCirc(180,-40){8}
\Text(173.5,-39)[c]{\tiny \bf $H/A$}
\Text(200,-30)[c]{ \bf $\widetilde{\chi}_{i}^0$}
\Text(150,-30)[c]{ \bf $\widetilde{\chi}_{j}^0$}
\ArrowLine(250,-50)(280,-50)
\LongArrow(250,-50)(220,-50)
\Vertex(250,-50){2}
\ArrowLine(110,-50)(80,-50)
\LongArrow(110,-50)(140,-50)
\Vertex(110,-50){2}
\LongArrow(280,-60)(210,-60)
\LongArrow(280,-60)(300,-60)
\Vertex(280,-60){2}
\LongArrow(80,-60)(150,-60)
\LongArrow(80,-60)(60,-60)
\Vertex(80,-60){2}
\Text(295,-60)[c]{ \bf $\widetilde{\chi}_{1}^0$}
\Text(50,-60)[c]{ \bf $\widetilde{\chi}_{1}^0$}
\Text(255,-40)[c]{ \bf $\widetilde{l_1}$}
\Text(90,-40)[c]{ \bf $\widetilde{l_2}$}
\Text(205,-50)[c]{ \bf $l_1$}
\Text(140,-50)[c]{ \bf $l_2$}
\Text(153,-60)[c]{ \bf $l'_1$}
\Text(193,-60)[c]{ \bf $l'_2$}
\Line(250,-40)(250,-50)
\Line(110,-40)(110,-50)
\Line(280,-50)(280,-60)
\Line(80,-50)(80,-60)
\Text(30,-50)[c]{ \bf (c)}
\end{picture} \
}
\caption{\it Angular configuration for maximizing the four-lepton invariant mass in the limit where the Higgs mass is very large (a), or for lower Higgs' masses with  $r_{1s} ~<~r_{2s}$ (b)
or $r_{1s} ~>~r_{2s}$ (c). }
 \label{m4lmaxfig}
\end{figure}

 To illustrate this phenomenon, consider $M_{4l}$ with $i=j=2$  and
 $\widetilde{l_2} = \widetilde{l_1} \equiv \widetilde{l}$ for simplicity.
Depending on the hierarchy of the ratios
\begin{equation}\label{ratios}
  r_{2A} \equiv \left(\frac{m_{\widetilde{\chi}_{2}^0}}{m_A}  \right)^2~,  ~~~~
  r_{1s} \equiv \frac{\left(\frac{m_{\widetilde{\chi}_{1}^0}}{m_{\widetilde{l}}}  \right)^2}
  {\left(1+\left(\frac{m_{\widetilde{\chi}_{1}^0}}{m_{\widetilde{l}}} \right)^2\right)^2 }~, ~~~~
  r_{2s} \equiv  \frac{\left(\frac{m_{\widetilde{l}}}{m_{\widetilde{\chi}_{2}^0}}  \right)^2}
  {\left(1+\left(\frac{m_{\widetilde{l}}}{m_{\widetilde{\chi}_{2}^0}} \right)^2\right)^2 }
\end{equation}
we find that one of the following three expressions is the endpoint:
 \begin{eqnarray}
   M_{4l}\left[ --++ \right] &\equiv &
   \frac{(m_{\widetilde{\chi}_{2}^0}^2 - m_{\widetilde{\chi}_{1}^0}^2)
   ( \Delta + m_A^2 )
   }
   {2 m_A m_{\widetilde{\chi}_{2}^0}^2}
    \\  \nonumber
   M_{4l}\left[ ++-- \right] & \equiv&
      \frac{m_{\widetilde{\chi}_{1}^0}^2 m_{\widetilde{\chi}_{2}^0}^2
      (\Delta + m_A^2) -
      m_{\widetilde{l}}^4 \Delta
      + m_A^2 (m_{\widetilde{l}}^4 - 2 m_{\widetilde{l}}^2 m_{\widetilde{\chi}_{2}^0}^2 )}
       {2 m_A m_{\widetilde{\chi}_{2}^0}^2 m_{\widetilde{l}}^2 }\\  \nonumber
   M_{4l}\left[ ---- \right]  & \equiv &
   \frac{m_{\widetilde{\chi}_{1}^0}^2 m_{\widetilde{\chi}_{2}^0}^2
      ( \Delta - m_A^2 ) -
      m_{\widetilde{l}}^4 \Delta
       -m_A^2 (m_{\widetilde{l}}^4 - 2 m_{\widetilde{l}}^2 m_{\widetilde{\chi}_{2}^0}^2 )}
       {2 m_A m_{\widetilde{\chi}_{2}^0}^2 m_{\widetilde{l}}^2 }
    \\  \nonumber
    {\rm with}~~\Delta & \equiv & \sqrt{m_A^4 - 4 m_A^2 m_{\widetilde{\chi}_{2}^0}^2} \\ \nonumber
 \end{eqnarray}
 \begin{table}
 \caption{\small \emph{Correct endpoint expressions for the four-lepton invariant mass $M_{4l}$ for all possible orderings of the ratios defined in (\ref{ratios}). }}
    \begin{center}
     \begin{tabular}{|l|l|} \hline
   Hierarchy & Endpoint  \\ \hline
  $r_{2A}~<~r_{1s},r_{2s}   $
           &  $[--++]$ \\ \hline
  $ r_{1s} ~<~ r_{2A}~<~r_{2s}   $
          & $[++--]$   \\ \hline
  $ r_{1s} ~<~ r_{2s} ~<~ r_{2A}   $
           & $[++--]$  \\ \hline
  $r_{2s} ~<~ r_{1s} ~<~ r_{2A}$
          & $[----]$ \\ \hline
  $ r_{2s} ~<~ r_{2A}~<~r_{1s}   $
          & $[----]$  \\ \hline
       \end{tabular}
    \end{center}
 \label{tab:endpt}
\end{table}   where the notation suggests the values of
 $\cos{\theta_i},~\cos{\theta_j},~\cos{\theta_1},$ and $\cos{\theta_2}$ which equal
 1 ('+') or -1 ('-'). Each of the six possible orderings of the ratios in (\ref{ratios}) single out the largest of these endpoints, as summarized in Table \ref{tab:endpt}.
 Thus, kinematic configurations that maximize $M_{4l}$ change as some masses become larger relative to others. In the particular limit where $m_A$ is much larger than the other masses, the hierarchy listed in the first row of Table  \ref{tab:endpt} applies with the angular configuration of Fig.~\ref{m4lmaxfig}a; this makes physical sense since the leptons should like to maximize their energies by being emitted parallel to their highly energetic mother neutralinos. As the Higgs mass is lowered, however, it may maximize total leptonic energy to emit one or both leptons antiparallel (cf.
 Fig.~\ref{m4lmaxfig}b,c) depending on the precise ratios (\ref{ratios}).

A similar situation applies to the other invariants in (\ref{avinv}): there exist three different endpoints depending on mass ratios, though the physical interpretation is not as clear as for $M_{4l}$. We have collected analytical expressions for all such endpoints in the Appendix.\footnote{We have not derived an endpoint for $a_4$ because the tail of this distribution is extremely shallow and therefore unlikely to be useful in the MC analysis.}

The distributions of (\ref{asymm4}) and (\ref{avinv}) also contain information in their shapes and specifically their peaks. We have not  derived analytical formulae for these (though in principle this is possible via the method of Ref.\cite{inv-shapes}). Peak values of distributions can always be numerically computed and we will find these useful as additional constraints (this is in fact the only use we have found for $a_4$). A more detailed discussion of this strategy appears in the following sections.

\section{Monte Carlo Results}
\label{sec:mc}
 In this section we would like to investigate how well the programme sketched above works for Monte Carlo (MC) LHC events  simulated at actual points in MSSM parameter space.
 We employ  the HERWIG 6.5\cite{HERWIG65} MC package (whose
MSSM input information comes from ISASUSY\cite{ISAJET} through ISAWIG\cite{ISAWIG} and HDECAY\cite{HDECAY})\footnote{ The CTEQ 6M\cite{CTEQ6} set of parton
distribution functions is used with top and bottom quark masses
set to $m_t=175\, \hbox{GeV}$ and $m_b=4.25\, \hbox{GeV}$, respectively.}  to
generate LHC events for an integrated luminosity of $300\, \hbox{fb}^{-1}$,
roughly equivalent to several years' high luminosity data, which we then run through private programs simulating a typical LHC detector environment.

We pass $pp \to H/A$ events with four hard and isolated\footnote{Specifically $p_T^\ell > 10, 8\, \hbox{GeV}$
for $e^{\pm},\mu^{\pm}$, respectively; $|\eta^{\ell}|<2.4\,$; isolation requires
no tracks of other charged particles in a $r = 0.3\, \hbox{rad}$ cone
around the lepton, and less than $3\, \hbox{GeV}$ of energy deposited into
the electromagnetic calorimeter for $0.05\, \hbox{rad} < r < 0.3\,
\hbox{rad}$ around the lepton.} leptons with flavor structures
 $e^+e^- \mu^+\mu^- $,  $e^+e^- e^+e^- $, or  $\mu^+\mu^- \mu^+\mu^- $ (hereafter  designated the 'isolated-4l cut') at the following three parameter points:
 \begin{description}
  \item[MSSM Sample Point 1]
   \begin{eqnarray*}
    \mu = 410\, \hbox{GeV}  ~~ ~~ M_2 = 260\, \hbox{GeV} ~~ ~~\tan \beta = 20 \\
     M_{\widetilde{e, \mu}_{L,R}} = 150\, \hbox{GeV} ~~~~ ~~~~ M_{\tilde{\tau}_{L,R}} = 250\, \hbox{GeV}
      \\
  M_A = 600\, \hbox{GeV} ~~~~ ~~~~ M_{\widetilde{q,g}} = 900\, \hbox{GeV}
   \end{eqnarray*}
    \item[MSSM Sample Point 2]
     \begin{eqnarray*}
    \mu = 150\, \hbox{GeV}  ~~ ~~ M_2 = 380\, \hbox{GeV} ~~ ~~\tan \beta = 10 \\
     M_{\widetilde{e, \mu}_{L,R}} = 150\, \hbox{GeV} ~~~~ ~~~~ M_{\tilde{\tau}_{L,R}} = 250\, \hbox{GeV}
      \\
  M_A = 500\, \hbox{GeV} ~~~~ ~~~~ M_{\widetilde{q,g}} = 900\, \hbox{GeV}
   \end{eqnarray*}
  \item[Snowmass Benchmark SPS1a\cite{benches}]
    \begin{eqnarray*}
    \mu = 352\, \hbox{GeV}  ~~ ~~ M_2 = 193\, \hbox{GeV} ~~ ~~\tan \beta = 10 ~\\
     M_{\widetilde{e, \mu}_{L,R}} = 136\, \hbox{GeV} ~~~~ ~~~~ M_{\tilde{\tau}_{L,R}} = 135\, \hbox{GeV}
      \\
  M_A = 394\, \hbox{GeV} ~~~ M_{\tilde{q}} \approx 500\, \hbox{GeV} ~~~ M_{\tilde{g}} \approx 600\, \hbox{GeV}
   \end{eqnarray*}
\end{description}
with other relevant sparticle masses at these points appearing in Table \ref{tab:mass}.

These points are chosen because the dominant source of signal events is
$H/A \to \widetilde{\chi}_{i}^0 \widetilde{\chi}_{i}^0$ ($i=2$ for MSSM1 and SPS1a,
$i=3$ for MSSM2). This can in fact be deduced from kinematics alone as follows:
in Ref.\cite{EWwedgebox} it was demonstrated that all boxlike wedgebox plots with low jet activity follow predominantly from either Higgs or chargino pair decays; the dominating presence of the former over the latter can be estimated by the ratio of flavor-balanced to
flavor-unbalanced events, $ \mathcal{R}_\pm $, as explained in the Introduction.
Since all three of these points have a 'simple box' wedgebox plot with
 $ \mathcal{R}_\pm >> 1$, we could know prior to knowing any MSSM  parameters that the observed four-lepton events were mostly from the Higgs decays (\ref{hdecay}) with $i=j$ and degenerate slepton masses\footnote{We could also deduce that the sleptons are on-shell since off-shell decays have very different invariant mass distributions. See Appendix for discussion.}.

Before conducting a full investigation with backgrounds and further cuts, we first investigate the
signal $pp \to H/A$ channel only at Sample Point 1 so as to understand how well endpoints and peaks agree with their theoretical values.

\begin{table}
 \caption{\small \emph{Relevant physical masses  at the parameter points
 (all masses in GeV).}}
    \begin{center}
     \begin{tabular}{|c|c|c|c|c|c|c|c|} \hline
 &  ${\widetilde\chi}^0_1$
 &  ${\widetilde\chi}^0_2$
 &  ${\widetilde\chi}^0_3$
 &  ${\widetilde\chi}^0_4$
 &   $\widetilde{e_1}$, $\widetilde{\mu_1}$
 &  $A^0$
 &   $H^0$   \\  \hline
 MSSM 1 &  $127.7$  & $243.9$ & $415.3$ & $436.1$ & $156.2$ & $600.0$  & $600.4$ \\
  MSSM 2 &  $120.9$  & $158.1$ & $205.4$ & $399.9$ & $156.2$ & $500.0$  & $500.4$ \\
  SPS1a & $96.1$ & $176.8$ & $358.8$& $377.8$  & $143.0$ & $393.6$  & $394.0$ \\ \hline
       \end{tabular}
    \end{center}
 \label{tab:mass}
\end{table}

\begin{figure}[!htb]
\begin{center}
\includegraphics[width=1.5in]{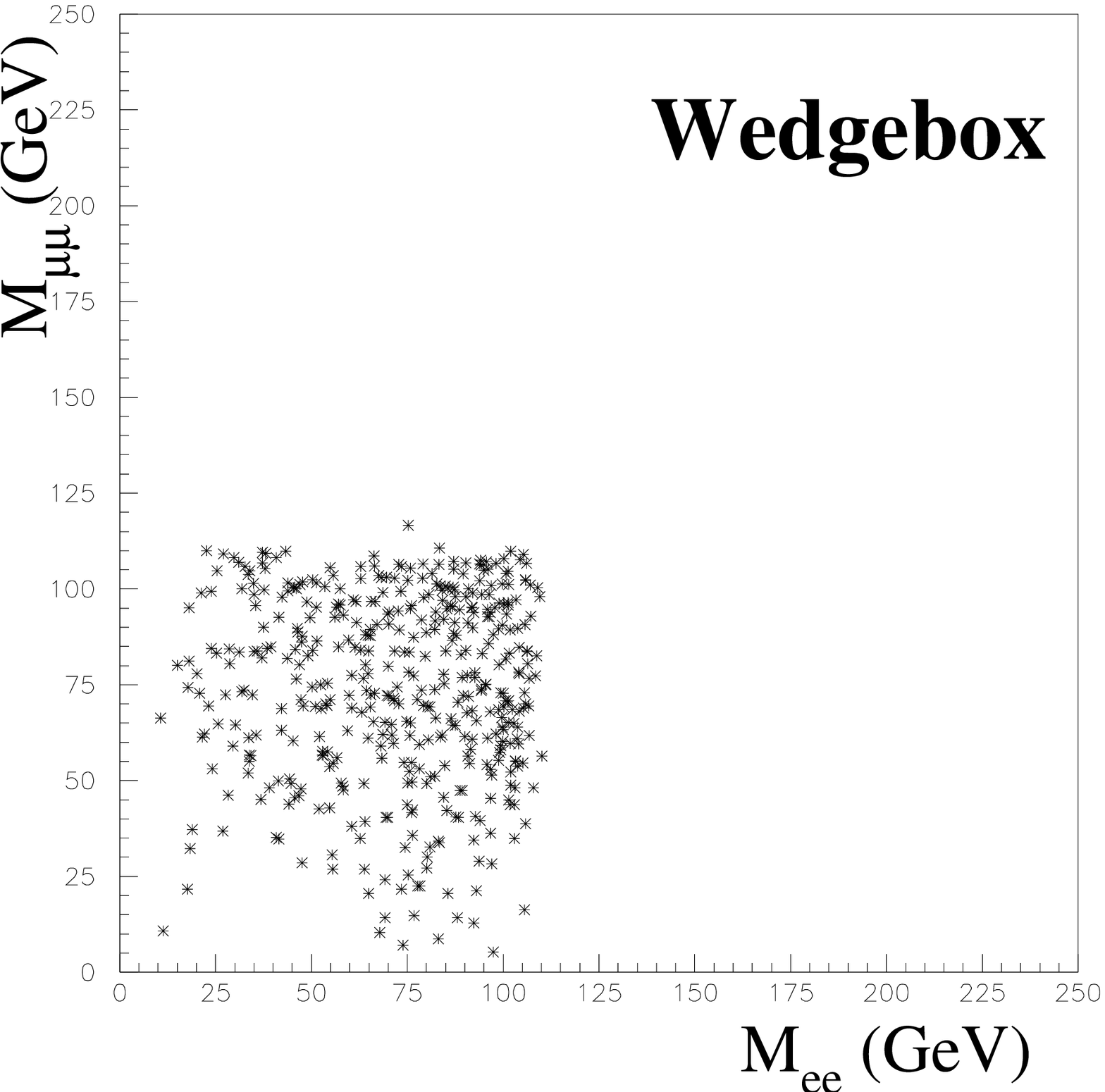}
\includegraphics[width=1.5in]{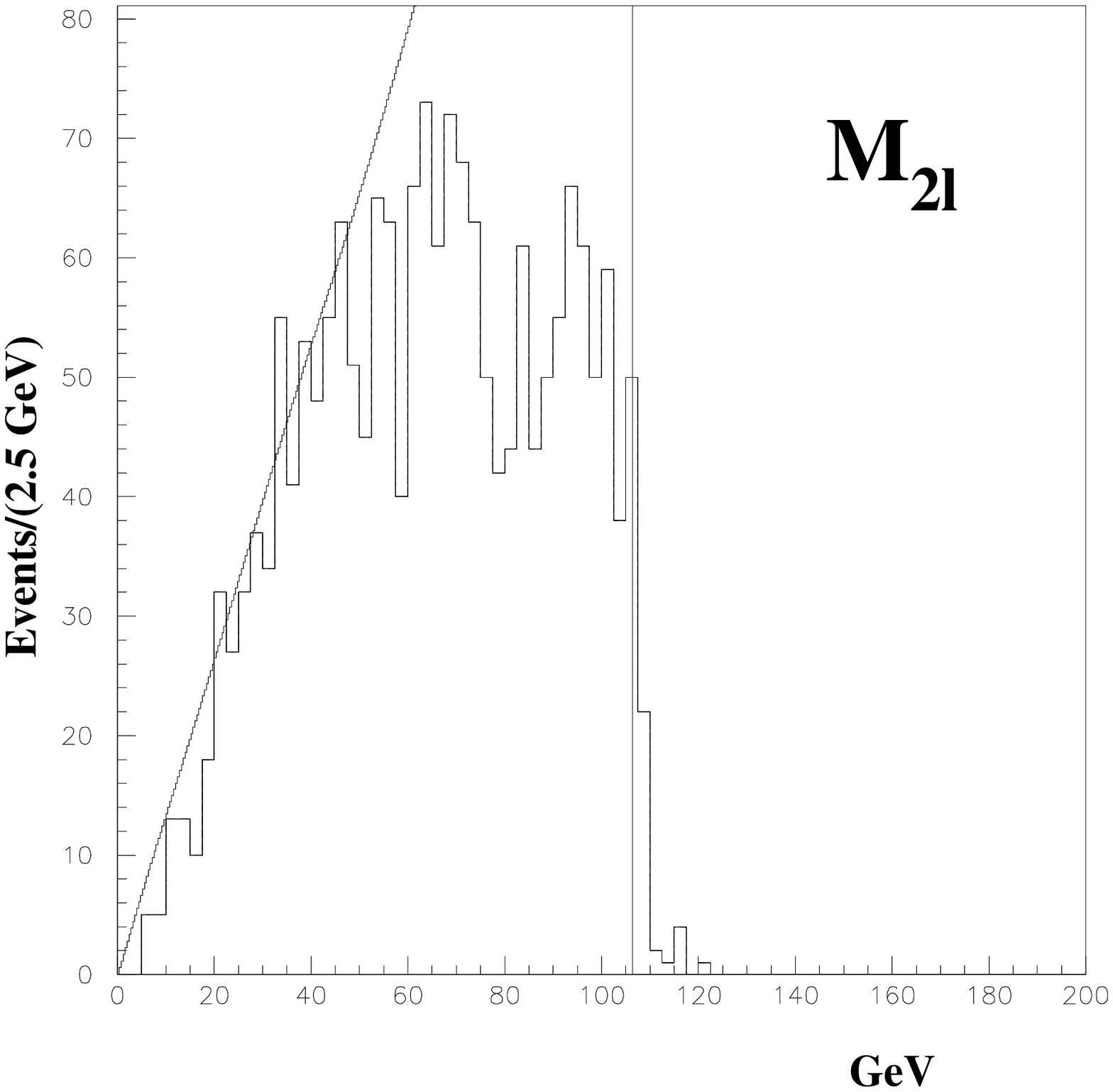}
\includegraphics[width=1.5in]{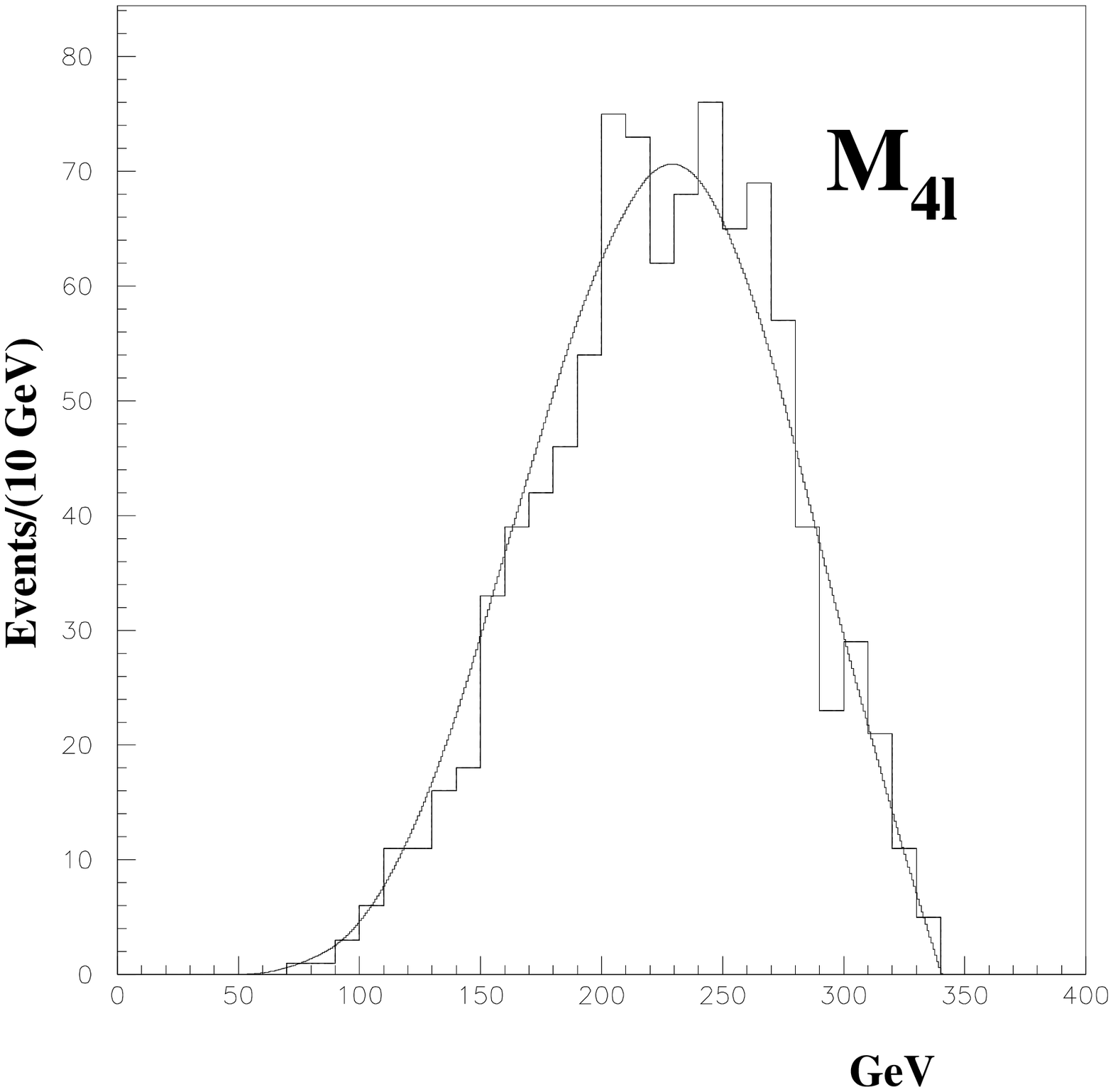}
\includegraphics[width=1.5in]{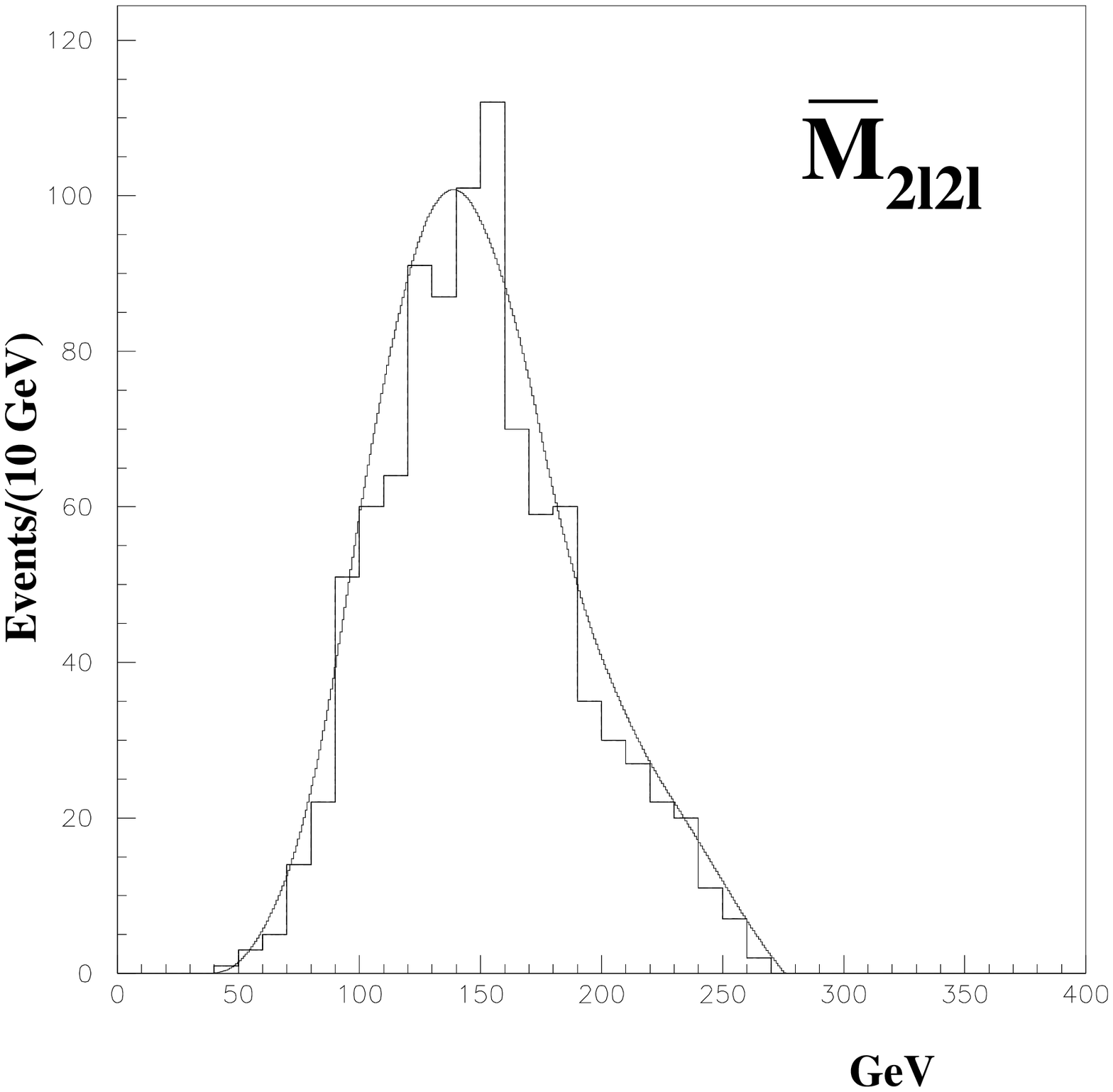}
\includegraphics[width=1.5in]{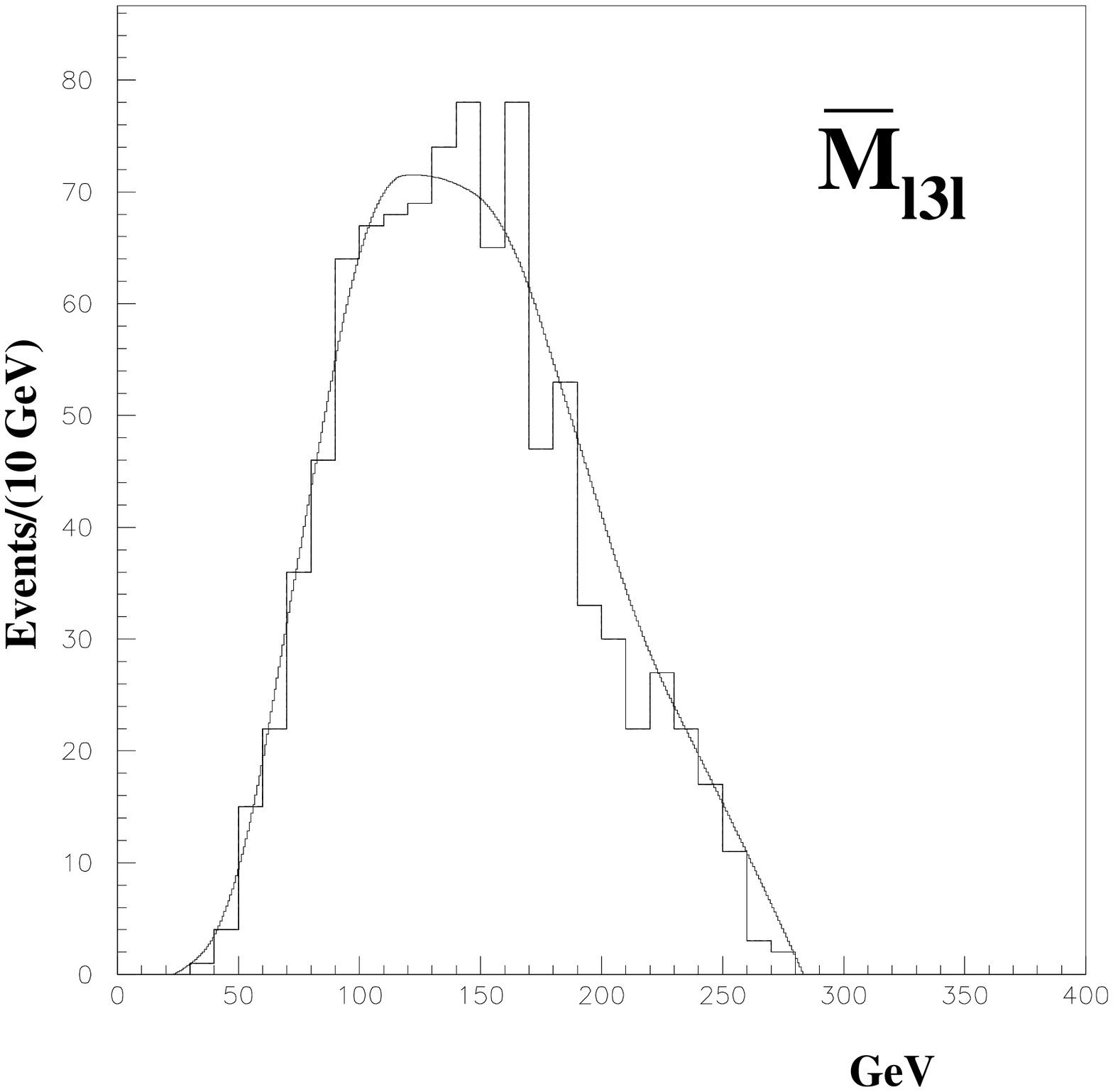}
\includegraphics[width=1.5in]{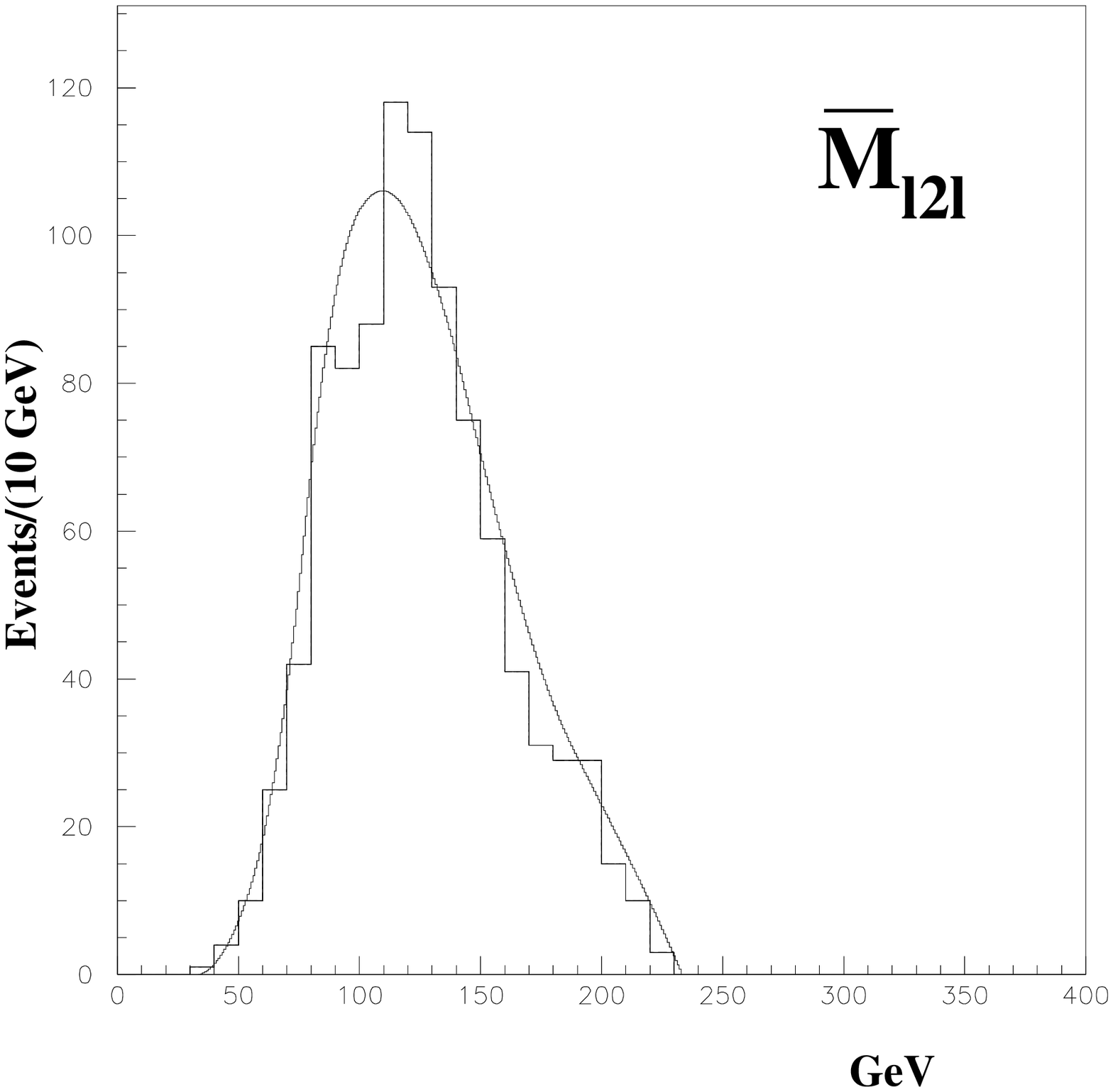}
\includegraphics[width=1.5in]{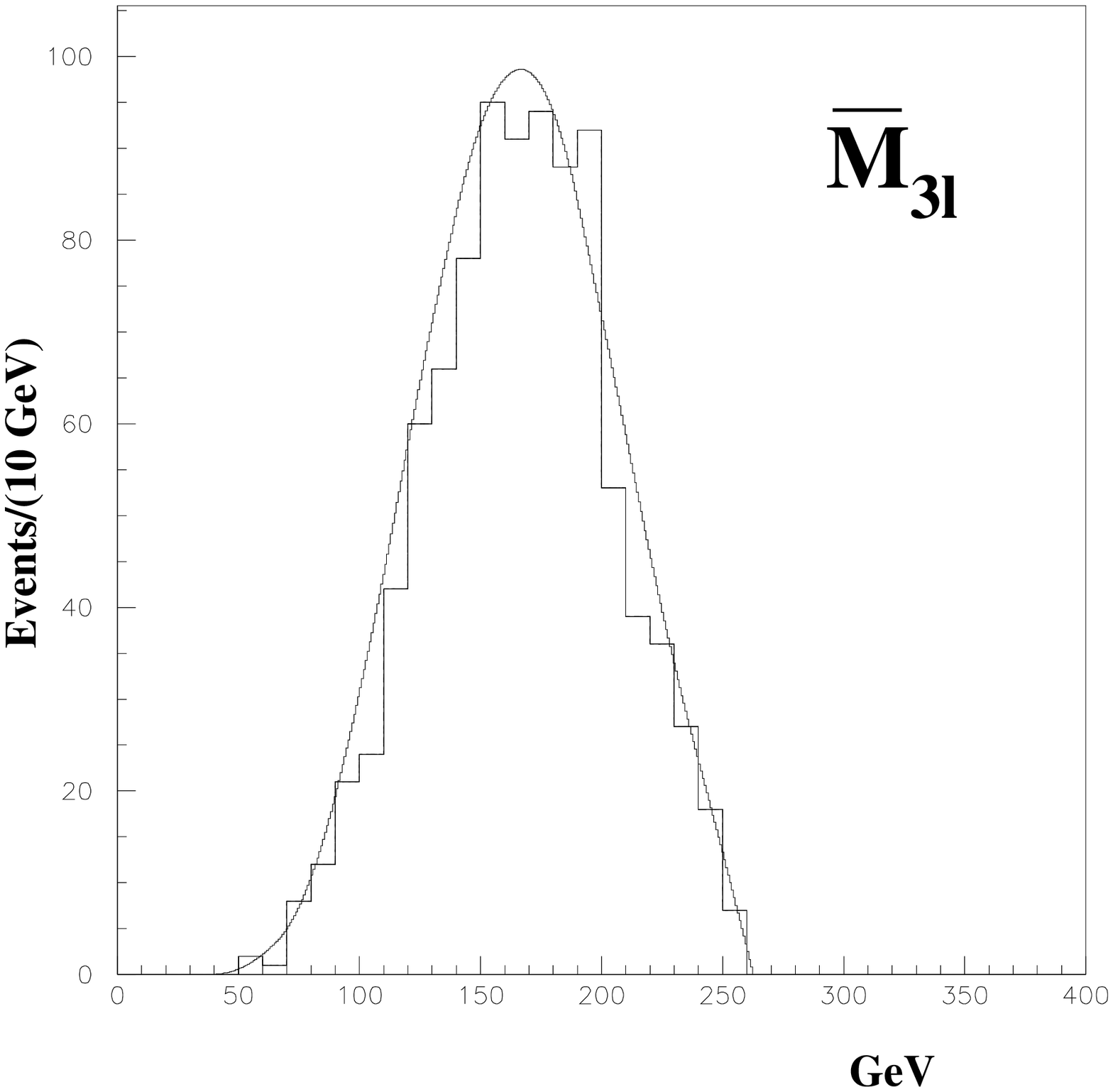}
\includegraphics[width=1.5in]{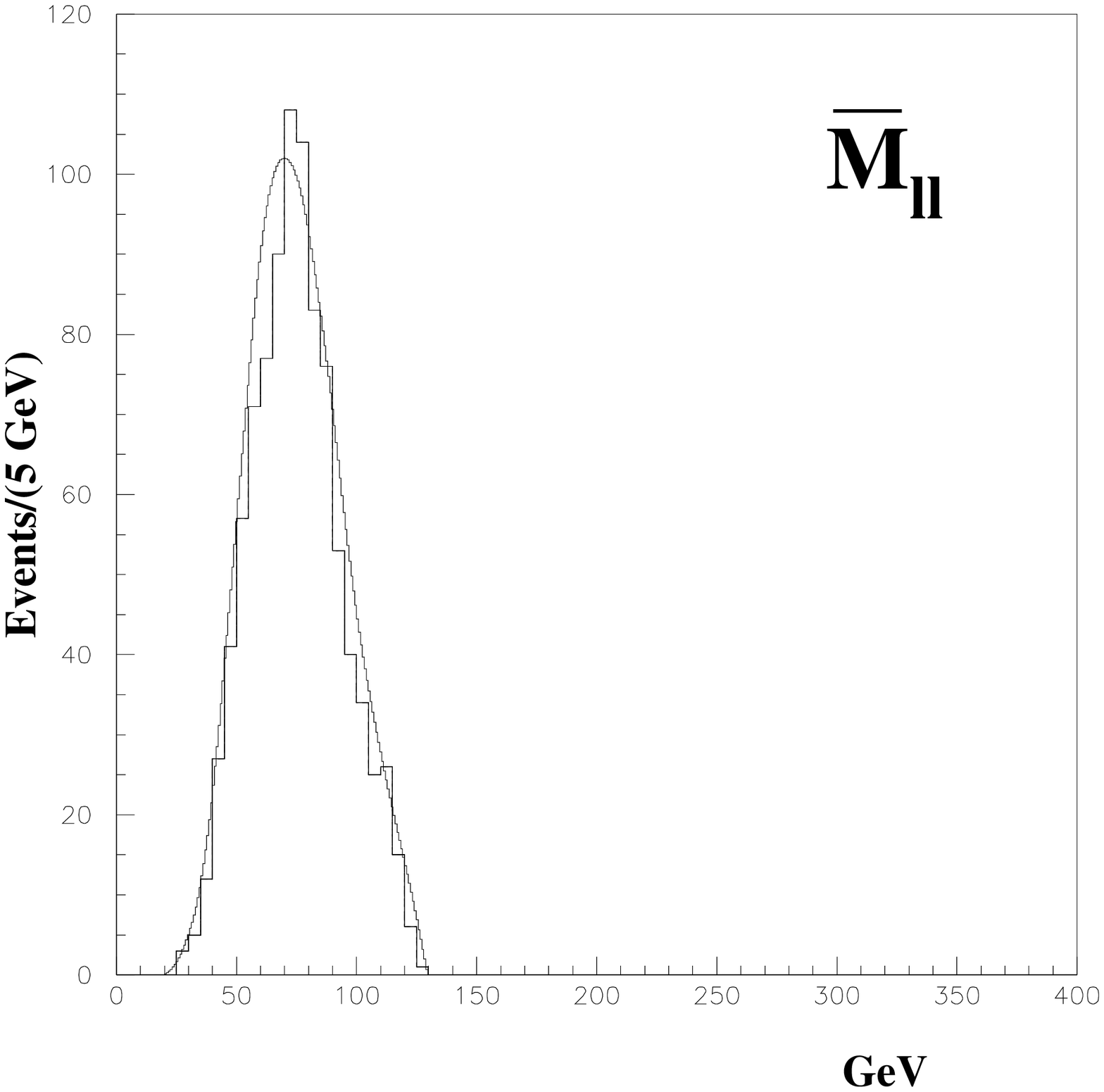}
\includegraphics[width=1.5in]{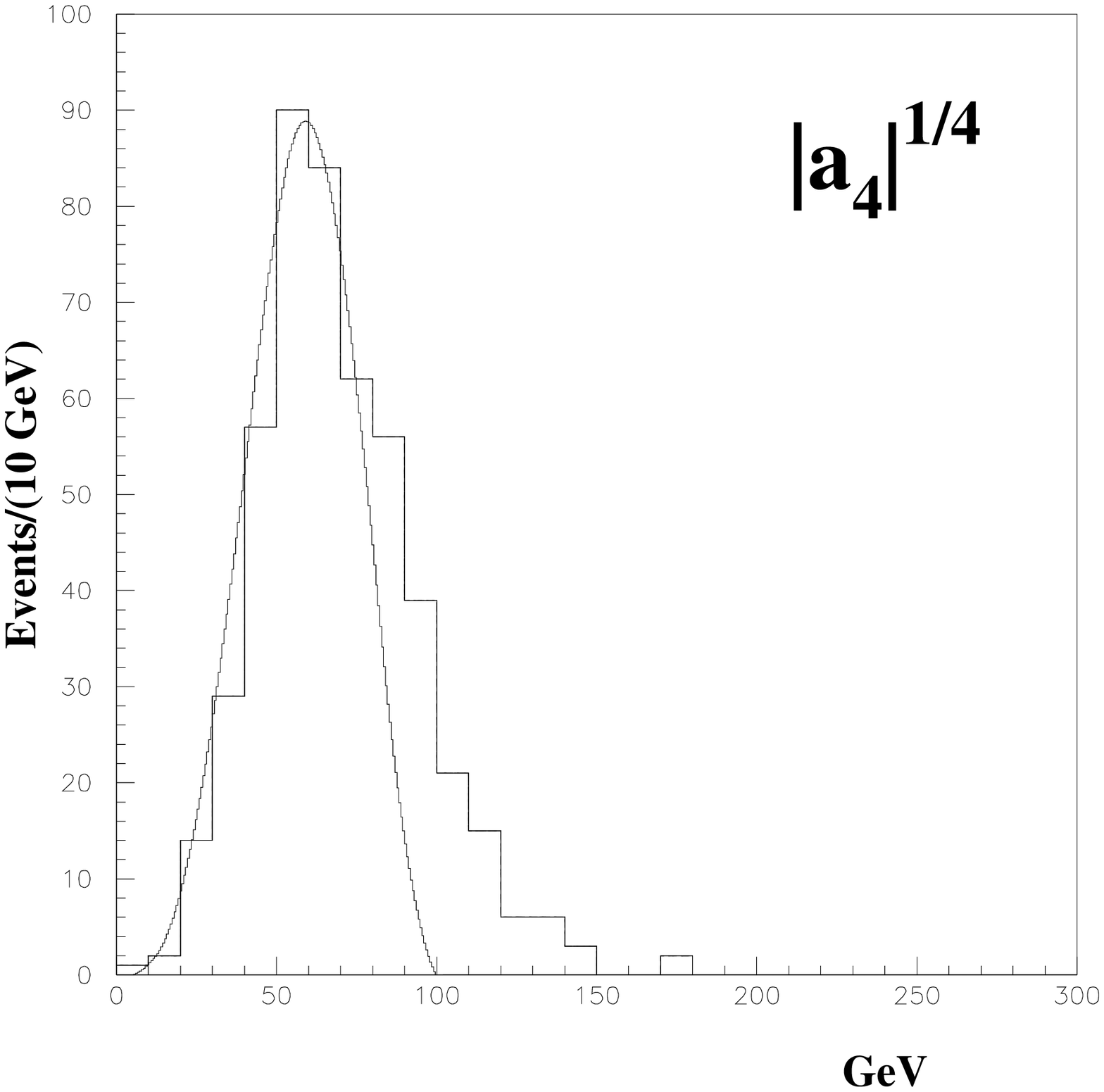}
\end{center}
\caption{\small \emph{Monte Carlo results for $H/A$ signal only at MSSM Sample Point 1 for $300\, \hbox{fb}^{-1}$,
with spline fits to theoretical distributions ($10^6$ events from kinematics only) overlaid and normalized (by signal rate/$10^6$).}
}
 \label{mc410sig}
\end{figure}

\subsection{H/A Signal Only}
The dominant source of lepton pairs at Sample Point 1 is
\begin{equation}\label{domdecay}
   pp \to H/A \to \widetilde{\chi}_{2}^0 \widetilde{\chi}_{2}^0 \to
    \widetilde{l_1}^\pm {l_1}^\mp \widetilde{l_2}^\pm {l_2}^\mp
    \to {l_1}^\mp {l'_1}^\pm {l_2}^\mp {l'_2}^\pm \widetilde{\chi}_{1}^0 \widetilde{\chi}_{1}^0
\end{equation}
For  $300\, \hbox{fb}^{-1}$ luminosity we obtain nearly 1000 $H/A$ events after the
isolated-4l cut; this rate is quite good but by no means optimized. With a lighter Higgs or higher $\tan\beta$ (increasing the Higgs production cross section) or heavier L-handed sleptons (reducing spoiler modes such as
   $\widetilde{\chi}_{2}^0 \to  \tilde{\nu} \nu$
   which compete with decays to $\tilde{e},\tilde{\mu}$), for example, the rate could increase by a factor of several\cite{EWwedgebox}.
   Distributions of all kinematic invariants are shown in Fig.~\ref{mc410sig}. We first notice that the wedgebox plot is a very crisp symmetrical box bounded by the $M_{2l}$ edge $\sim 108  \,\hbox{GeV}$ as expected.
   Other plots have overlying normalized fits to purely theoretical distributions which we have computed from relativistic kinematics only. We see a reasonable agreement between the two which,
of course, is not expected to be perfect since the theoretical distributions do not include the
 isolated-4l cut, detector effects, particle widths, and other features of the MC.
 Presumably such account for
 the suppressed peak of $M_{2l}$ and elongated tail of $|a_4|^{1/4}$, neither of which we will employ in any fits for this work.

Peaks and tails of MC distributions are generally well-fit with Gaussian functions\footnote{ Specifically, we fit peaks to a  3-parameter function $a + b~exp[-(x-c)^2]$ (a parabola would work just as well). Although tails happen to look quite linear at Sample Point 1, we have found that this is not generally true. There is always a slight amount of nonlinearity which, for economy's sake, we fit to the same functional form as used for the peaks.}. However, theoretical peaks may occasionally (as in the case of $\overline{M}_{l3l}$ in Fig.~\ref{mc410sig}) be somewhat asymmetric which gives rise to a slight ($2-5 \,\hbox{GeV}$)  error in comparison to the MC fits, deemed tolerable at this level of analysis.
We fit all histograms interactively in PAW, identifying three sources of error: binning, fit interval, and statistical error on the parameters of the fit; these are added in quadrature. The small error in fitting theory peaks then arises almost totally from fit interval error. For MC data, endpoints are dominated by binning and fit interval, and peaks by fit parameter error.
 The agreement between fits for the signal and theory is very good (cf. Table \ref{tab:compare}) and stable when subject to a liberal range of missing energy and jet cuts (Table \ref{tab:cuts}), the use of which we will find necessary when backgrounds are included.

\begin{table}
 \caption{\small \emph{Comparison of Endpoints (EP) and Peaks (P) of Monte Carlo (MC) and theoretical values (boldface, based on relativistic kinematics only), for {\bf signal only} at MSSM Sample Point 1 (all numbers to GeV precision). Error is quadrature sum of bin size, fit interval, and fit parameter error.
Error on theory peaks arises from using a Gaussian to fit a slightly non-Gaussian shape.
 {\small (*)The endpoint for $|a_4|^{1/4}$ is very shallow and hard to fit, so the
 corresponding entry is blank.
 (**)We do not fit the peak for $M_{2l}$ which theoretically coincides with its endpoint.}
 }} \label{tab:compare}
    \begin{center}
     \begin{tabular}{|r|c|c|c|c|c|c|c|c|} \hline
    &  $M_{2l}$ &  $M_{4l}$ & $\overline{M}_{2l2l}$ &  $\overline{M}_{l3l}$ &   $\overline{M}_{l2l}$ &
    $\overline{M}_{3l}$ &   $\overline{M}_{ll}$ &  $|a_4|^{1/4}$ \\ \hline
    EP &  $105 \pm 5$  &  $340 \pm 10$  &  $275 \pm 10$  &  $285 \pm 10$  &  $230 \pm 10$  &   $262 \pm 10$  &  $130 \pm 10$  & * \\
        &   \textbf{108}   &  \textbf{345} &  \textbf{282} &  \textbf{288} & \textbf{235} &  \textbf{266} &  \textbf{131} & * \\ \hline
    P &  **   &  $228 \pm 17$  &  $142 \pm 10$  &   $131 \pm 15$ &  $116 \pm 10$ & $167 \pm 11$   &  $72 \pm 5$  &  $63 \pm 10$ \\
        &    **   &  $\mathbf{230 \pm 2}$     & $\mathbf{142 \pm 2}$     &   $\mathbf{132 \pm 5}$    &  $\mathbf{108 \pm 3}$     &  $\mathbf{168 \pm 2}$     &   $\mathbf{72 \pm 1}$    &   $\mathbf{60 \pm 1}$\\
     \hline
    \end{tabular}
    \end{center}
\end{table}

\begin{table}
 \caption{\small \emph{Effects of  missing energy ($ \slashchar{E_{T}}$)
 and jet ($ E_j$) cuts on fitted values of peaks and endpoints ($P|EP$) for the signal only at MSSM Point 1  (all units are GeV).  Variation of peak and endpoint values are seen to be about $\pm 5\, \hbox{GeV}$ and $\pm 3\, \hbox{GeV}$, respectively. This is smaller than the fitting error on these.
 }}  \label{tab:cuts}
    \begin{center}
     \begin{tabular}{|c|c|c|c|c|c|c|c|} \hline
          $( \slashchar{E_{T}}, E_j)$   & $M_{4l}$ &$\overline{M}_{2l2l}$ &$\overline{M}_{l3l}$ &$\overline{M}_{l2l}$ &$\overline{M}_{3l}$ & $\overline{M}_{ll}$ & $|a_4|^{1/4}$\\ \hline
  $(0,0)$ & $241 ~|~ 345$ & $150~|~ 272$& $124~|~275$ & $108~|~ 227 $ & $170~|~ 264$ &
  $72 ~|~ 128 $ & $ 66 ~|~ -$ \\ \hline
   $(0,30)$ & $244~|~ 347$ & $152 ~|~ 272$& $125 ~|~ 274$ & $114 ~|~ 227 $ & $173~|~ 264$ & $74~|~ 128 $ & $ 64 ~|~ -$ \\ \hline
    $(0,50)$ & $240 ~|~ 344$ & $151 ~|~ 273$& $134 ~|~ 277$ & $117~|~ 228 $ & $169 ~|~ 262$ &
  $76~|~ 129 $ & $ 68 ~|~ -$ \\ \hline
     $(0,100)$ & $241~|~ 342$ & $152 ~|~ 278$& $137 ~|~ 276$ & $115~|~ 232 $ & $173 ~|~ 265$ &
  $75 ~|~ 128 $ & $ 69 ~|~ -$ \\ \hline
     $(10,0)$ & $238 ~|~ 345$ & $154 ~|~ 271$& $132 ~|~ 275$ & $119 ~|~ 228 $ & $175 ~|~ 262$ &
  $74 ~|~ 131 $ & $ 68 ~|~ -$ \\ \hline
    $(10,50)$ & $242 ~|~ 343$ & $148 ~|~ 278$& $132 ~|~ 275$ & $117~|~ 227 $ & $174 ~|~ 264$ &
  $75 ~|~ 128 $ & $ 66 ~|~ -$ \\ \hline
     $(10,100)$ & $239 ~|~ 345$ & $152 ~|~ 272$& $138 ~|~280$ & $118~|~ 236 $ & $177 ~|~ 266$ &
  $74 ~|~ 127 $ & $65 ~|~ -$ \\ \hline
  $(30,0)$ & $231 ~|~ 344$ & $146 ~|~ 273$& $136 ~|~ 281$ & $115 ~|~ 229 $ & $171 ~|~ 264$ &
  $74 ~|~ 128 $ & $ 68 ~|~ -$ \\ \hline
    $(30,50)$ & $232~|~ 346$ & $149 ~|~ 270$& $136~|~ 278$ & $115~|~ 231 $ & $166 ~|~ 260$ &
  $75 ~|~ 129 $ & $61 ~|~ -$ \\ \hline
     $(30,100)$ & $233~|~ 346$ & $150 ~|~ 271$& $129 ~|~ 282$ & $116 ~|~ 230 $ & $175 ~|~ 264$ &
  $74 ~|~ 130 $ & $ 66 ~|~ -$ \\ \hline
                  \end{tabular}
    \end{center}
\end{table}

\subsection{With Backgrounds and Cuts}

Confident that a pure signal can be fit closely to the theory with a simple fitting procedure,
we now add all SM and MSSM backgrounds into our analysis.
SM backgrounds ($t \overline{t}, t b, t \overline{t} Z^0, W^+ W^-, Z^0 Z^0$) are eliminated\footnote{ Residual $pp \to Z^{(*)}Z^{(*)}$ ($i.e.$
either Z-boson may be on- or off-shell) backgrounds can be large (up to 1000 events after the isolated-4l cut) but are constant in MSSM space. In our analysis we generated six $Z^{(*)}Z^{(*)}$ backgrounds, using one subtracted from the average of the other five; cancellation was nearly total. Where this background, more-or-less forming two strips of events within $\Gamma_Z$ of $m_Z$ on the wedgebox plot, is external to the box structure, it can also be subtracted by inspection.} by the isolated-4l cut in addition to demanding at least $\slashchar{E_{T}} ~ > ~ 20\, \hbox{GeV}$ missing energy in each event.

The main SUSY backgrounds at Sample Points 1 and 2 are gaugino
processes $(\tilde{q},\tilde{g})\widetilde{\chi}_{i}^{0,\pm}$,
$\widetilde{\chi}_{i}^0 \widetilde{\chi}_{j}^0$, and $\widetilde{\chi}_{i}^\pm \widetilde{\chi}_{j}^\mp$,
 plus a handful of events from  $\tilde{l}$,$\widetilde{\nu}$, $t\overline{t}h$, and $tH^-$ (the reader may refer to our earlier works \cite{EWwedgebox,cascade} for details on all backgrounds and cut efficiencies for this four-lepton signal).
Backgrounds with colored sparticles can be suppressed via a jet cut:
only events with jets\footnote{Jets are defined by a cone algorithm with
$r=0.4$ and must have $|\eta^j|<2.4\,$ .} having reconstructed energy $E_j ~ < ~ 50\, \hbox{GeV}$ are allowed to pass.
Chargino and slepton backgrounds are greatly reduced by subtracting distributions of flavor-unbalanced events.

\begin{figure}[!htb]
\begin{center}
\includegraphics[width=2.0in]{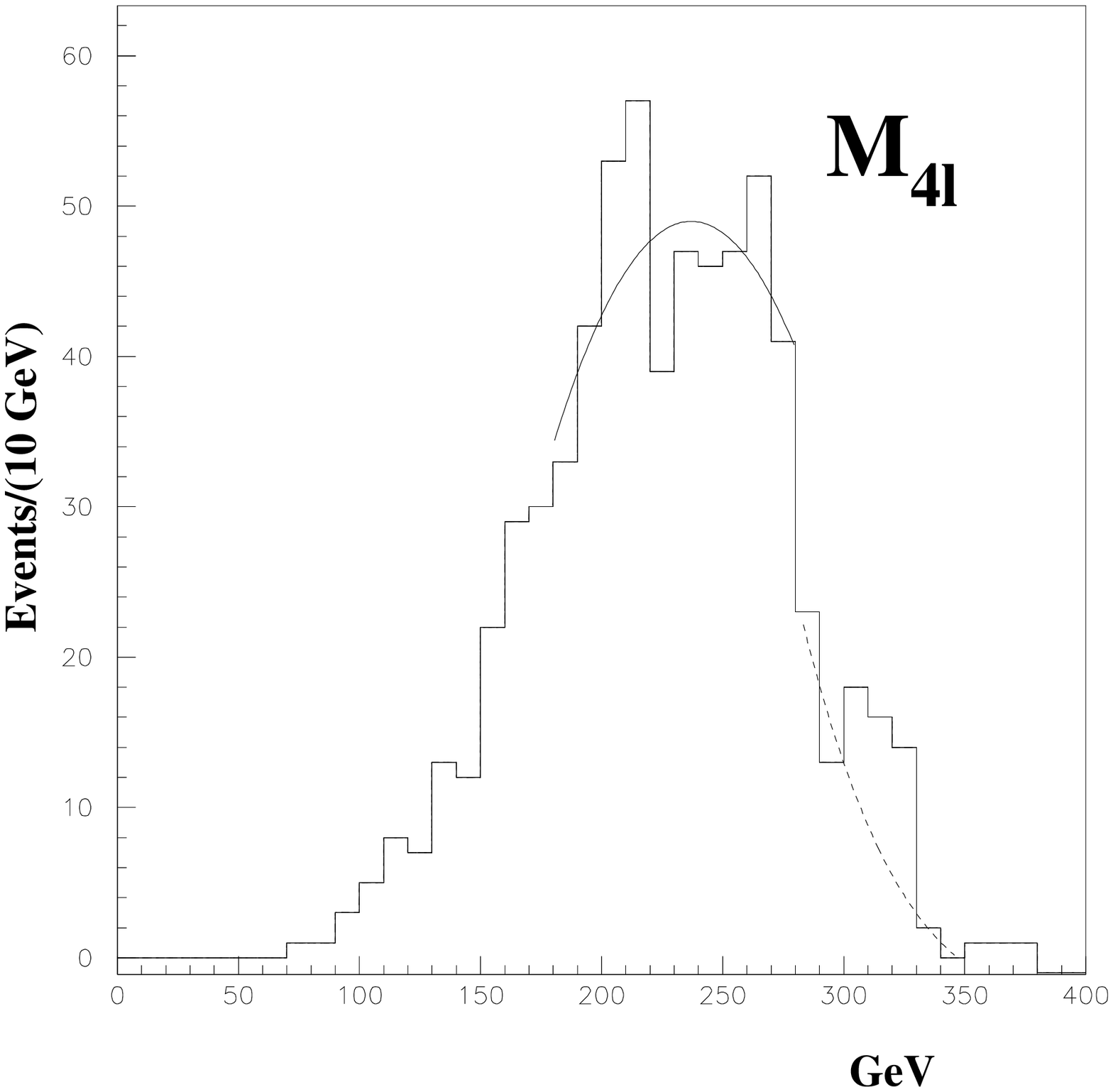}
\includegraphics[width=2.0in]{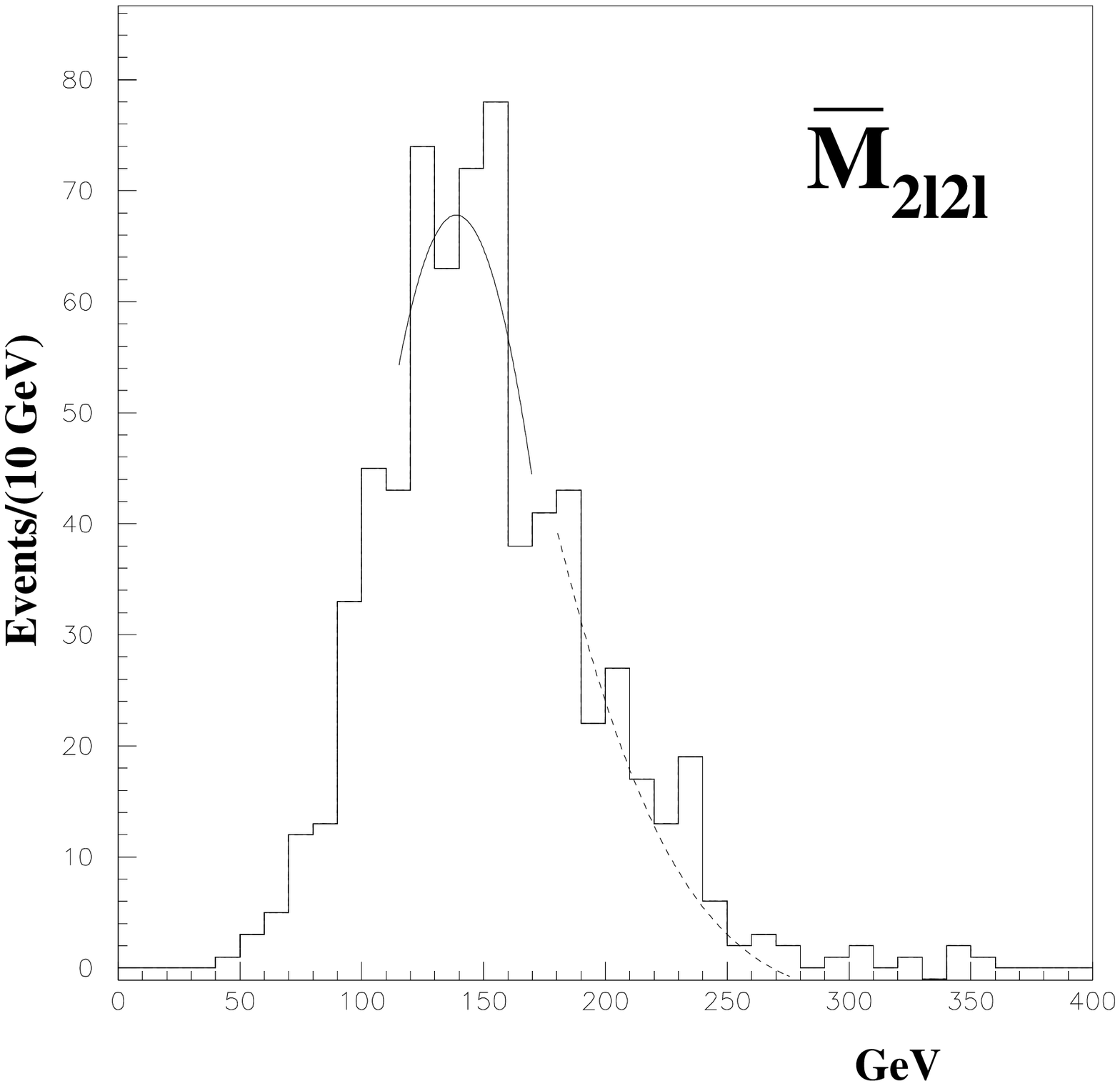}
\includegraphics[width=2.0in]{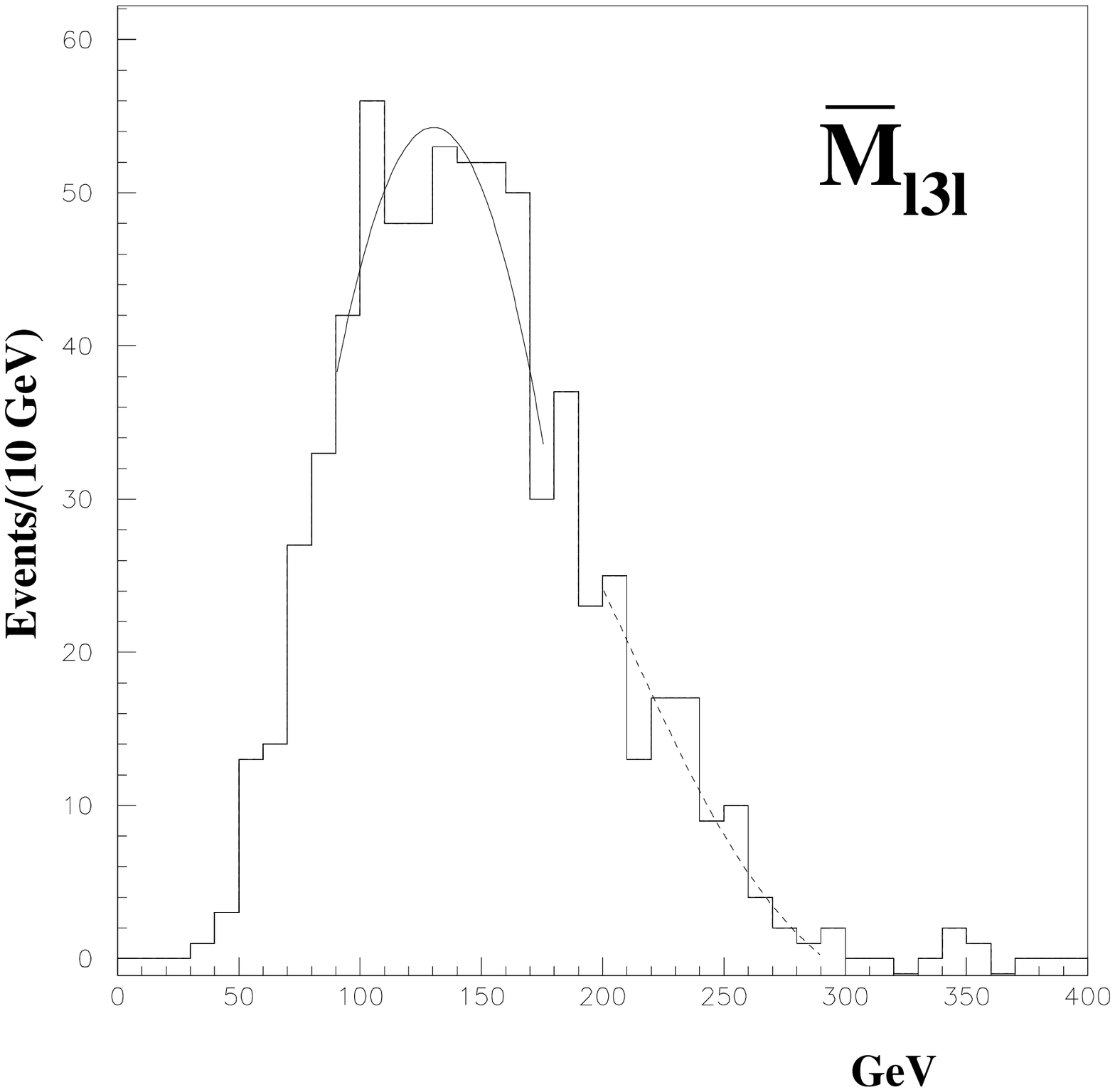}
\includegraphics[width=2.0in]{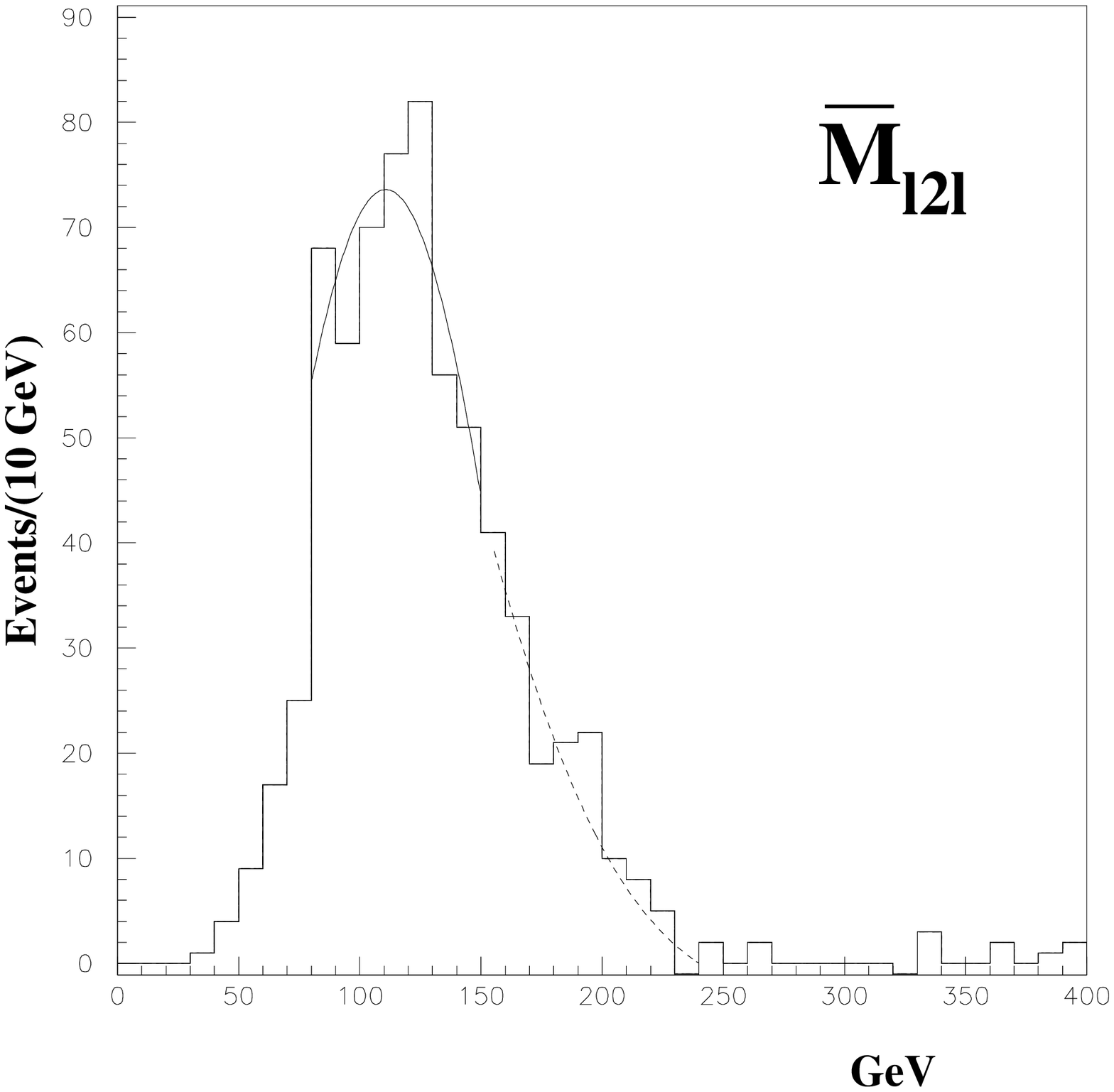}
\includegraphics[width=2.0in]{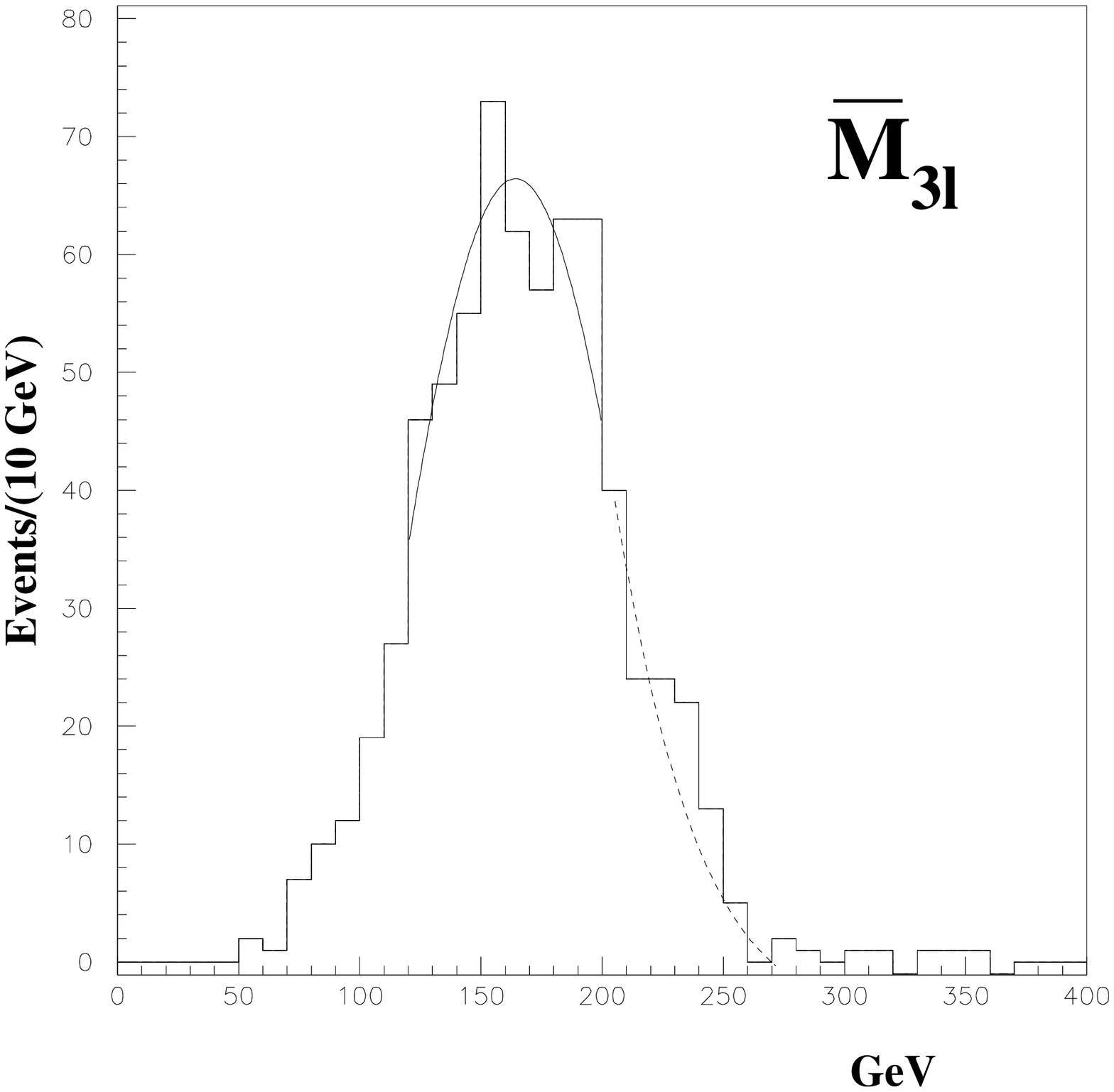}
\includegraphics[width=2.0in]{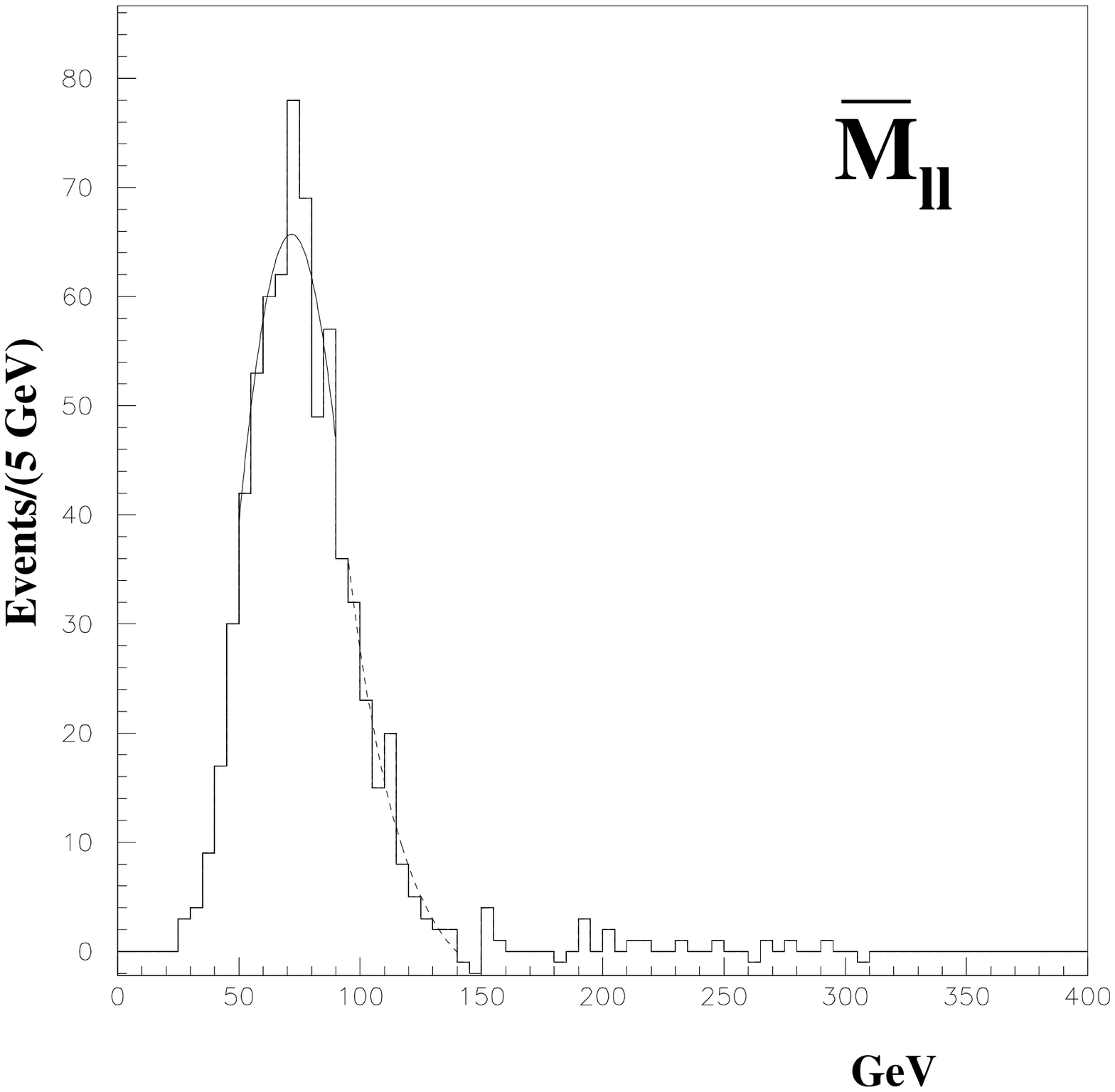}
\end{center}
\caption{\small \emph{ Monte Carlo results (for $300\, \hbox{fb}^{-1}$) at MSSM Sample Point 1 for signal and all backgrounds with missing energy and jet energy cuts described in the text. Flavor unbalanced distributions have been subtracted in this and the following plots.}
}
 \label{mc410}
\end{figure}

\begin{figure}[!htb]
\begin{center}
\includegraphics[width=1.8in]{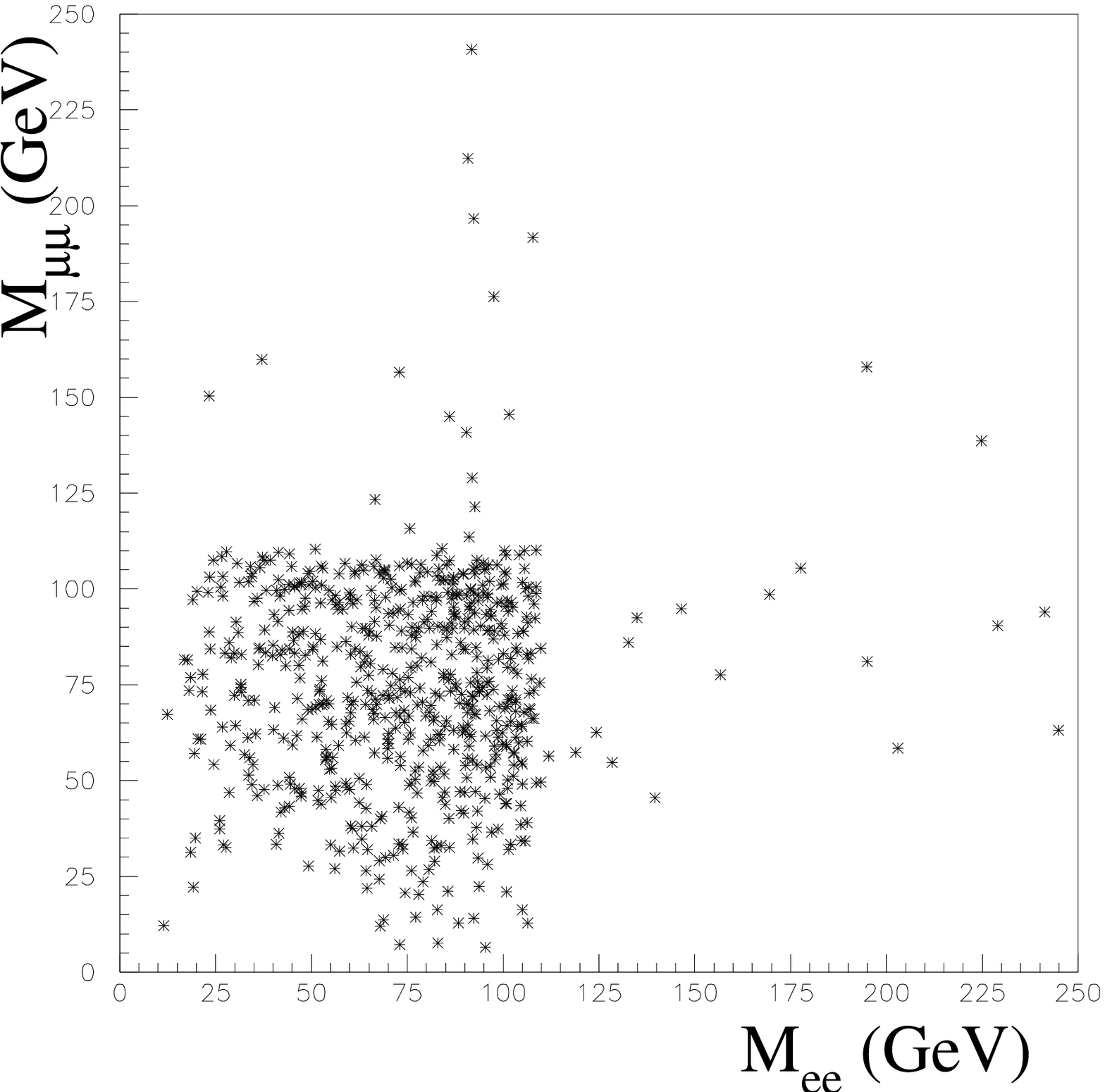}
\includegraphics[width=1.8in]{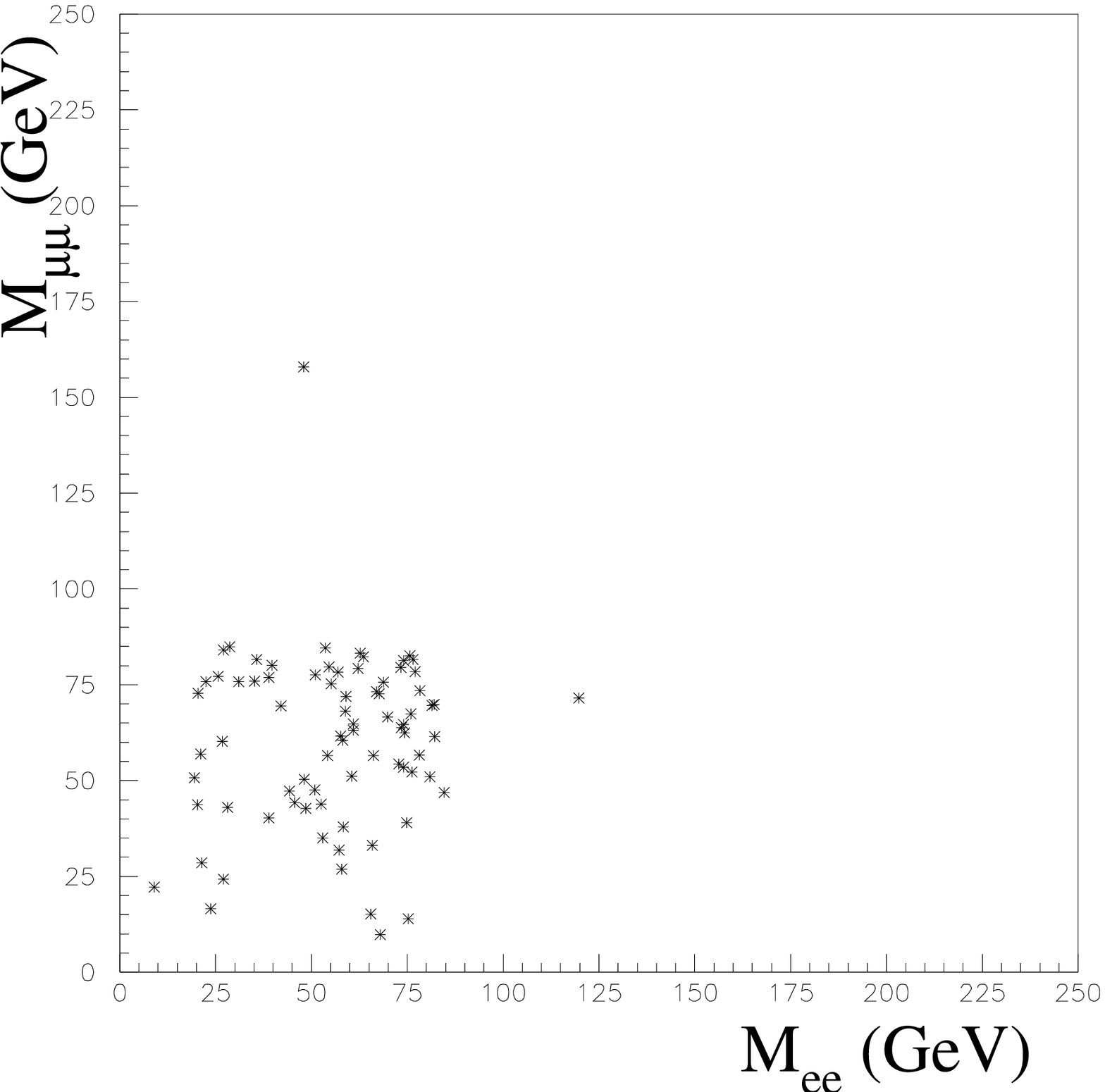}
\includegraphics[width=1.8in]{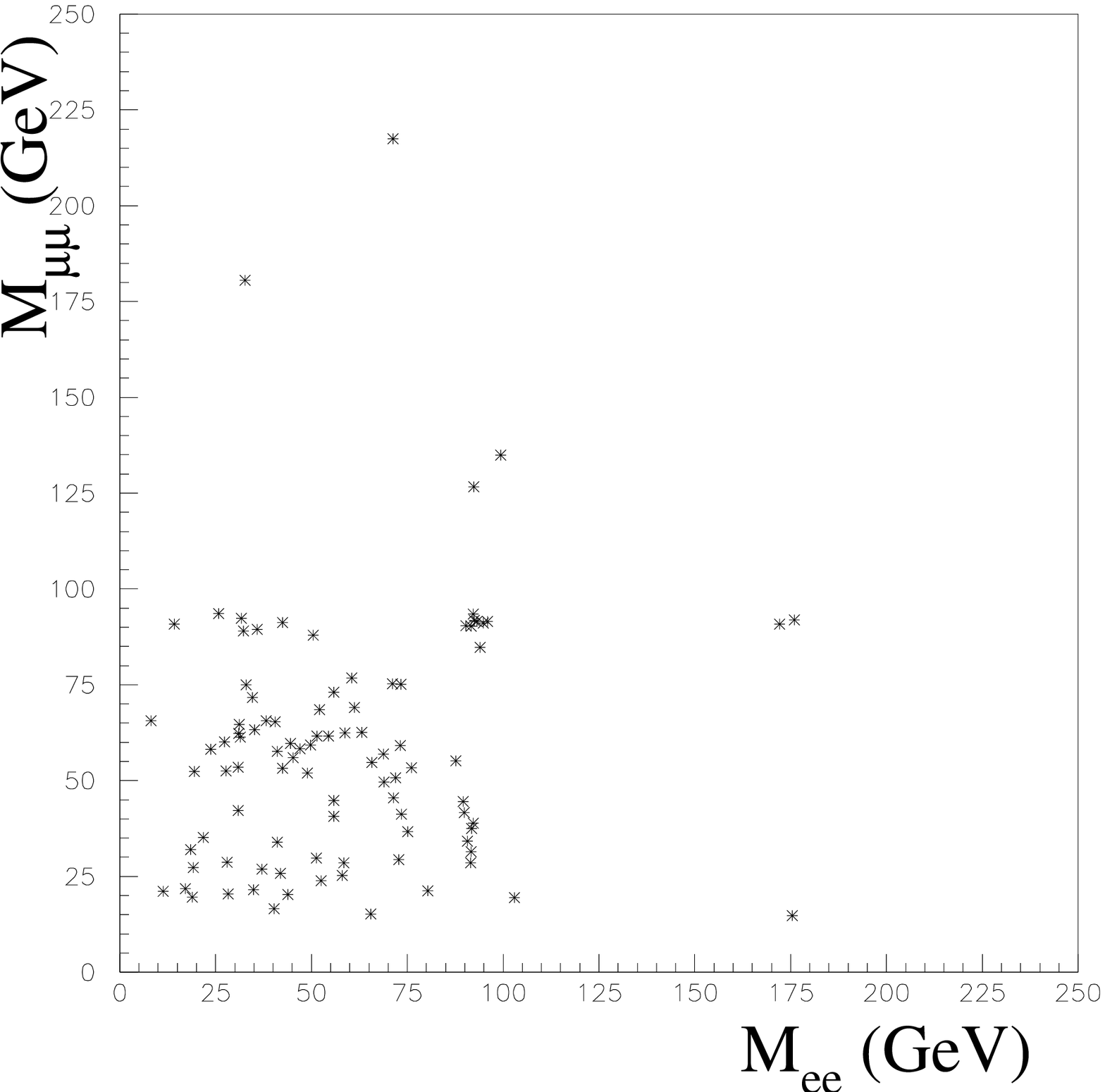}
\end{center}
\caption{\small \emph{ Wedgebox Plots (for $300\, \hbox{fb}^{-1}$) at MSSM Sample Point 1 (left), Sample Point 2 (center), and SPS1a (right) 
 for signal and all backgrounds after cuts.}}
 \label{mcwedge}
\end{figure}

In Fig.~\ref{mc410} we see the resulting distributions at Sample Point 1, which after cuts has a large value of $ \mathcal{R}_\pm = 786/48 \approx 16.3$.
 The wedgebox plot (Fig.~\ref{mcwedge}) retains a very dense box shape, though now with a few dozen background events scattered outside the perimeter. Peaks and endpoints  have drifted but they are still in good agreement with theoretical values (see Table \ref{tab:fits} for results at this and the other parameter points). The drift is mostly due to backgrounds, since the missing energy and jet cuts only have a mild ($\pm 5\, \hbox{GeV}$ or less) effect when these are varied across a liberal range (cf. Table \ref{tab:cuts}).

At Sample Point 2 the same backgrounds are more formidable, though here 
 $ \mathcal{R}_\pm =197/26 \approx 7.6$ is still large and we have
 a clear boxlike wedgebox topology in Fig.~\ref{mcwedge}. Therefore, by
 our hypothesis fitted values should be in reasonable agreement with those for a Higgs decay $H/A \to \widetilde{\chi}_{i}^0 \widetilde{\chi}_{i}^0$ (note, however, that this technique does not tell us which neutralino is involved\footnote{Here, in fact, $i=3$ since the smallness of the $H {\chi}_{2}^0 {\chi}_{2}^0$ coupling is more severe than phase-space suppression to the heavier $\widetilde{\chi}_{3}^0$-pair. Illustrating this interplay of parameters forms another reason for our choice of this MSSM parameter point.}).
As the shapes of the various distributions look rather similar to those at Sample Point 1, we do not display them. Lower statistics lead to larger fitting errors, though agreement with theory is nevertheless intact.

\begin{figure}[!htb]
\begin{center}
\includegraphics[width=2.0in]{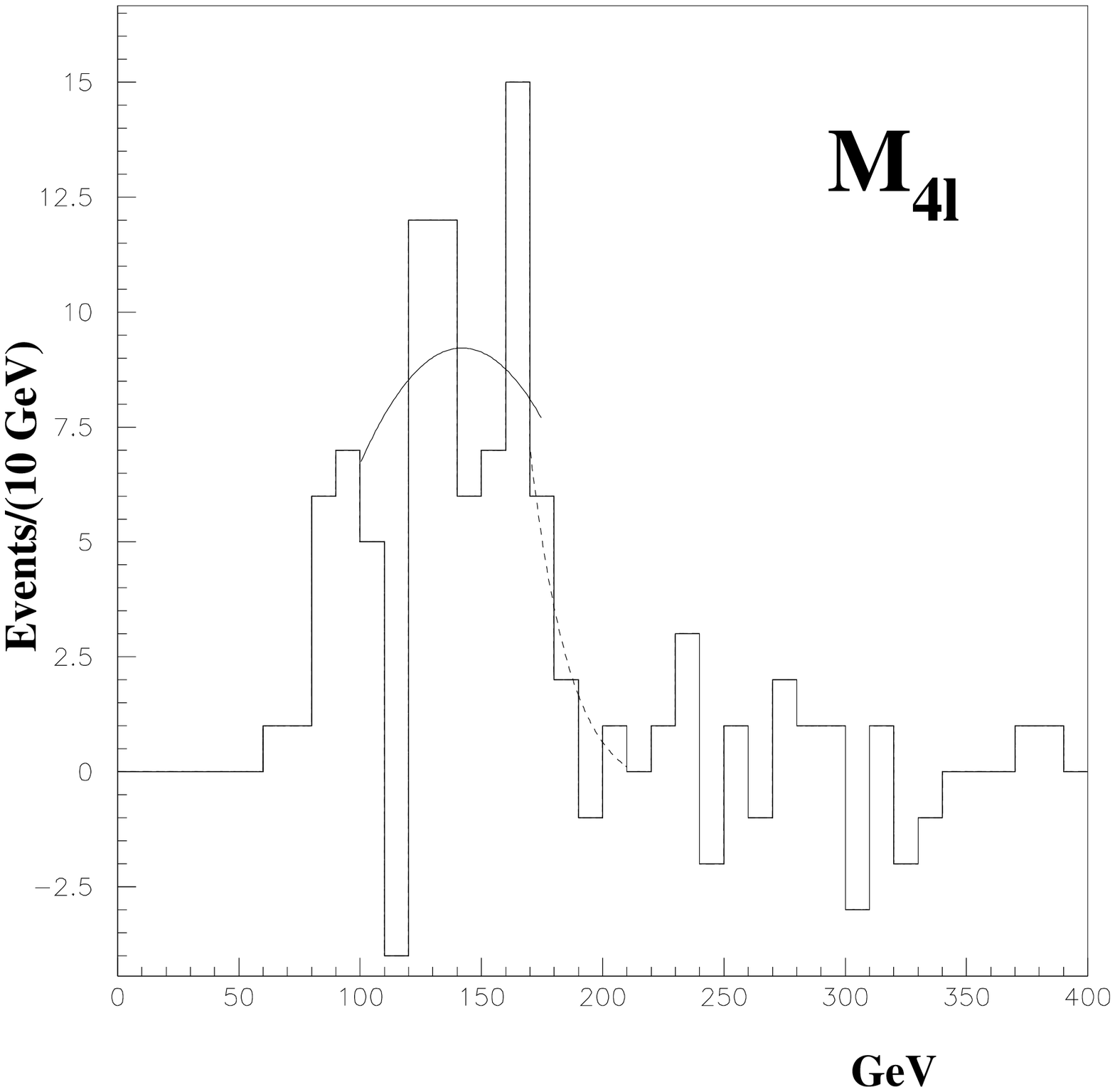}
\includegraphics[width=2.0in]{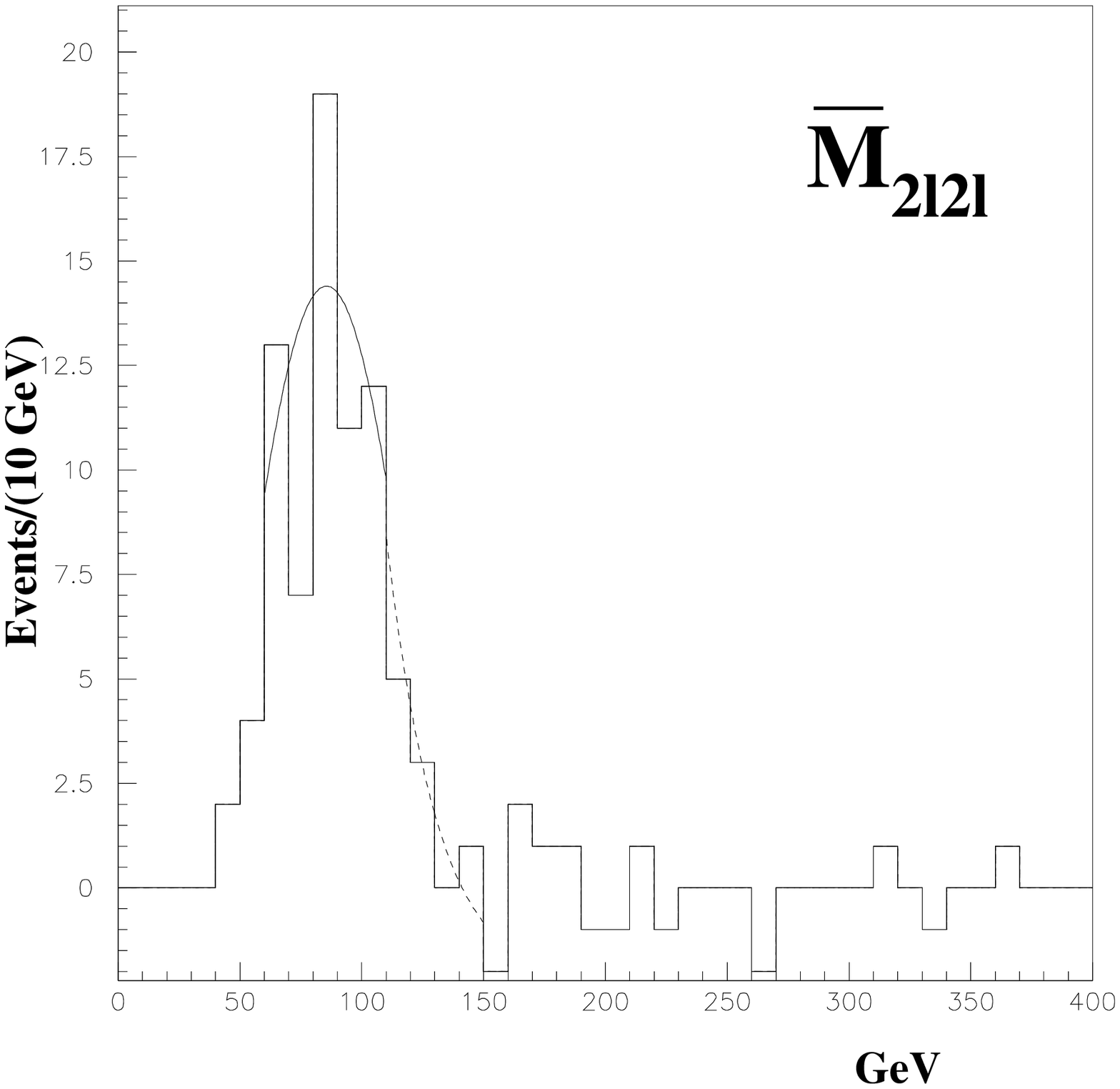}
\includegraphics[width=2.0in]{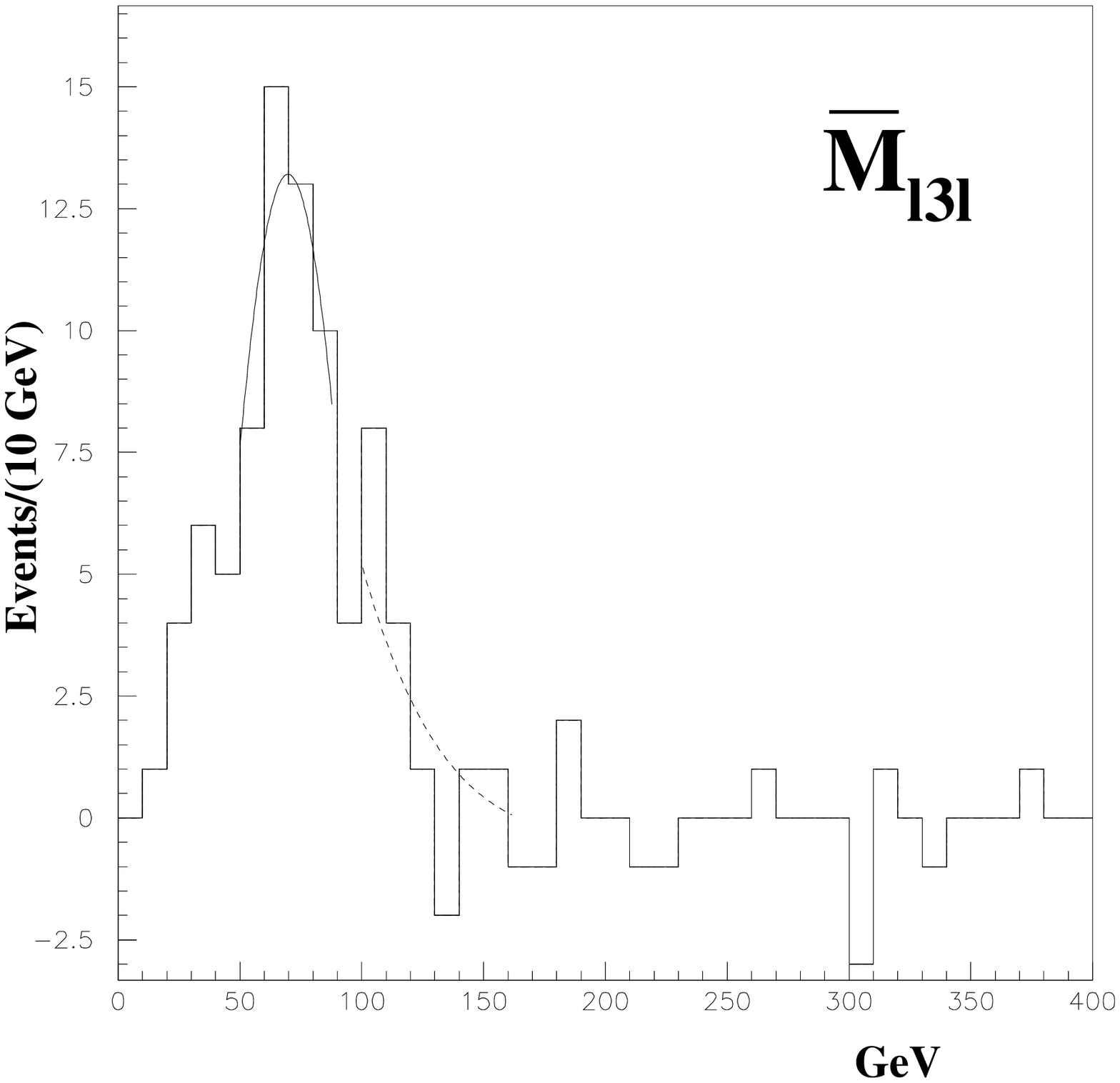}
\includegraphics[width=2.0in]{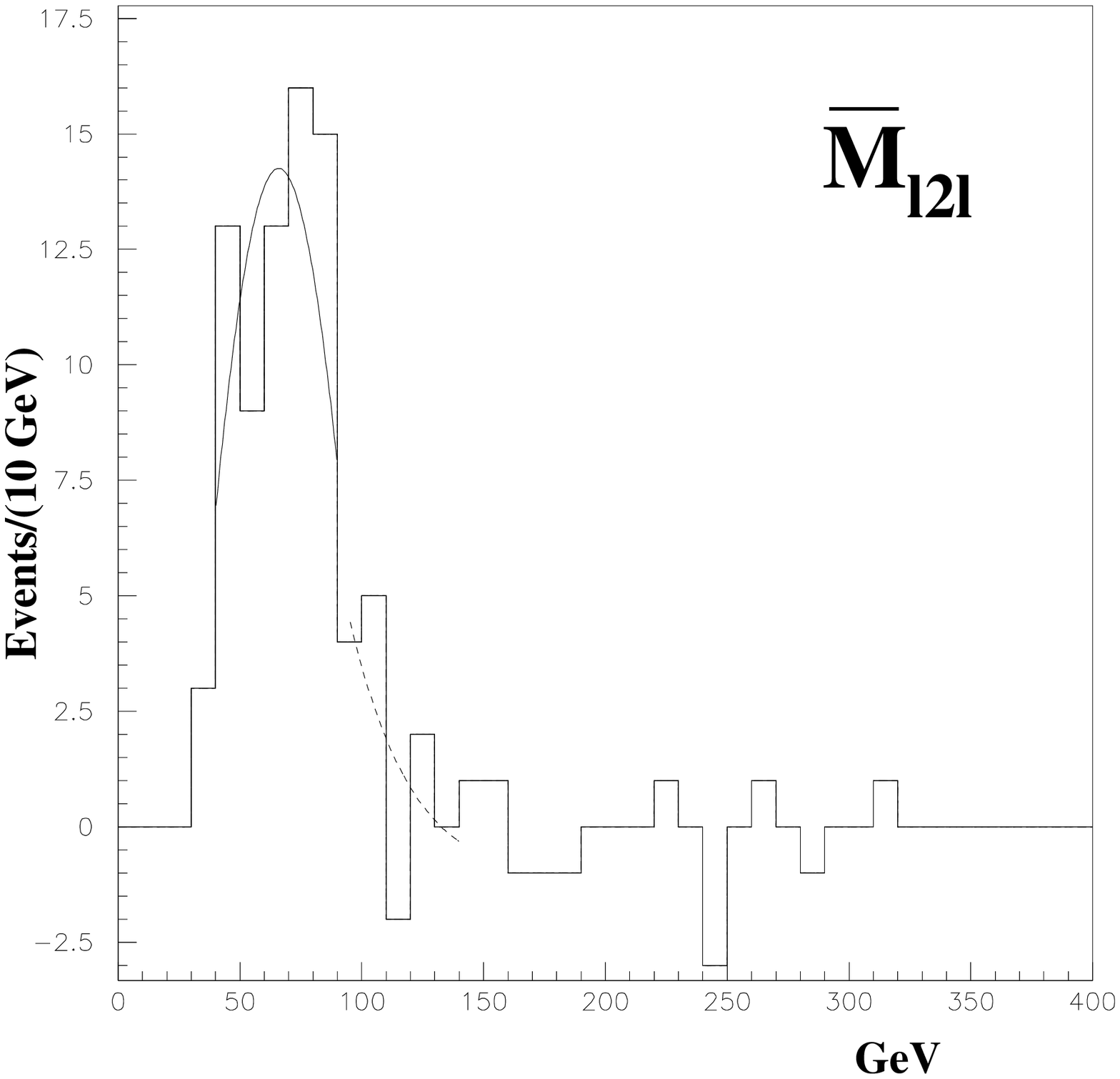}
\includegraphics[width=2.0in]{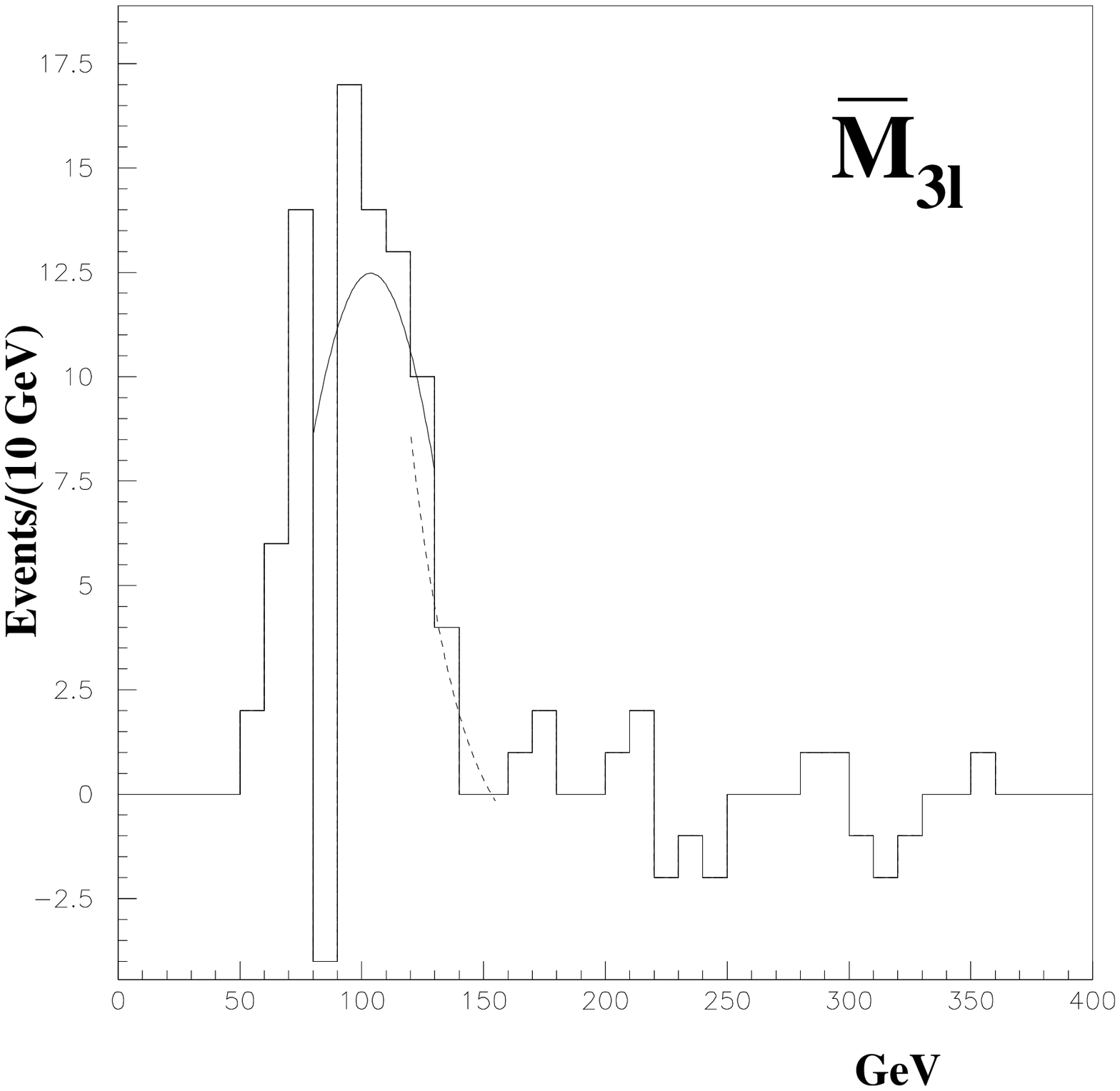}
\includegraphics[width=2.0in]{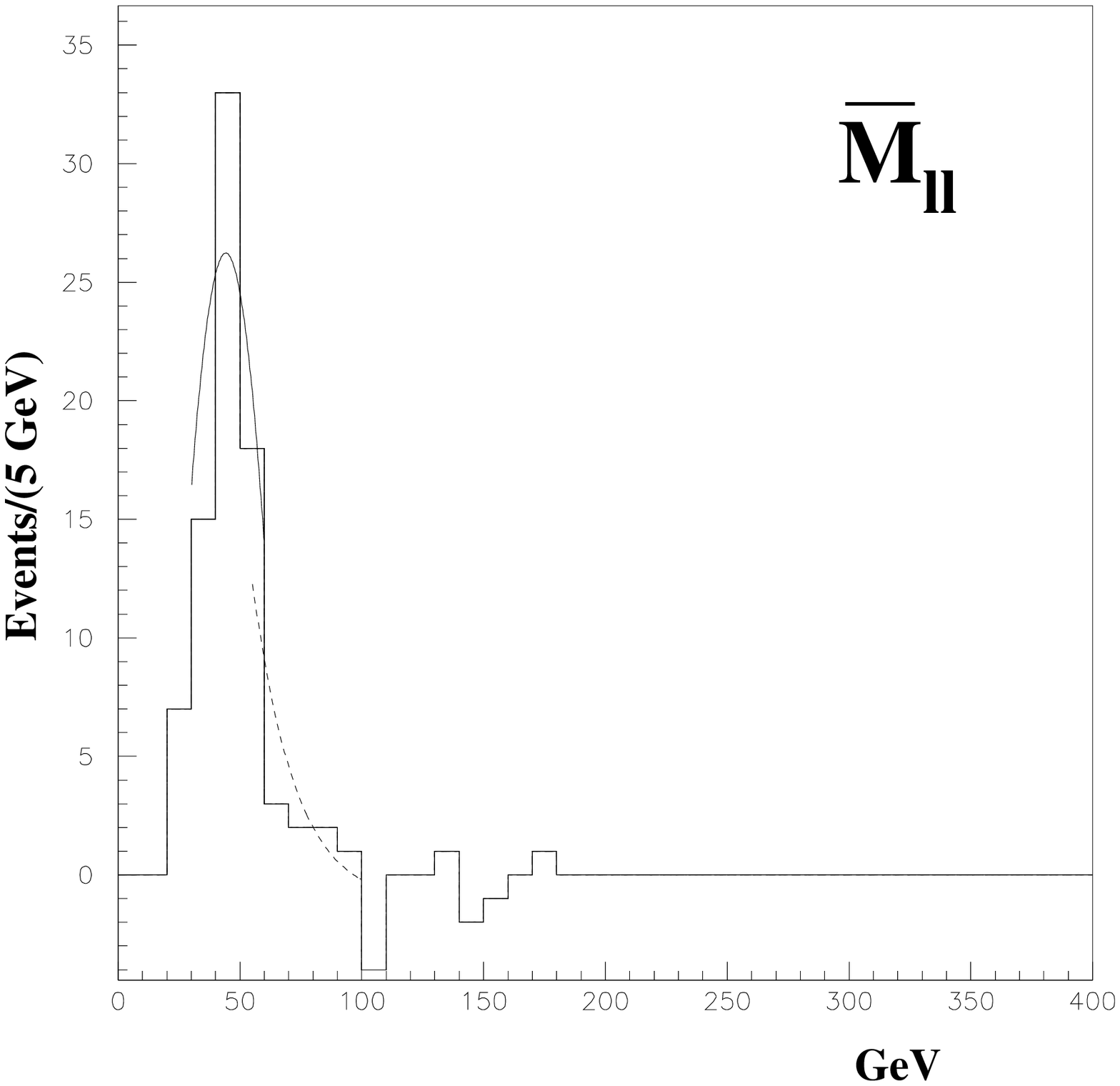}
\end{center}
\caption{ \small \textit{Monte Carlo results (for $300\, \hbox{fb}^{-1}$) at SPS1a for signal and all backgrounds with cuts described in the text.}}
 \label{mcsps}
\end{figure}

Snowmass Benchmark Point SPS1a would seem to present a more challenging case.
 With a lighter spectrum of colored sparticles, there arise significant backgrounds due to squarks and gluinos which force us to tighten the jet cut to  $E_j ~ < ~ 30\, \hbox{GeV}$; this however diminishes the signal ($H/A \to \widetilde{\chi}_{2}^0 \widetilde{\chi}_{2}^0$) close to the level of the irreducible SUSY backgrounds, giving a lower value of $\mathcal{R}_\pm =138/59 \approx 2.3$. Statistics now barely allow for the identification of a box topology in the wedgebox plot (Fig.~\ref{mcwedge}) and we would seem to be testing the limits of our technique in fitting the distributions in Fig.~\ref{mcsps}. Yet fitted central values are reasonably close to theoretical expectations in Table \ref{tab:fits} and still agree within error-bars.

\begin{table}
 \caption{\small \emph{Comparison of MC Endpoints (EP)  and Peaks (P) (all numbers to GeV precision)for MSSM Sample Points 1,2 and SPS1a to theoretical values in boldface. Theoretical endpoints are calculated from formulae in the Appendix, whereas all other values are fits as described in the text.
 }}  \label{tab:fits}
    \begin{center}
     \begin{tabular}{|c|c|c|c|c|c|c|} \hline
     & MSSM1:EP  &MSSM1:P& MSSM2:EP &MSSM2:P& SPS:EP &SPS:P\\    \hline
     $M_{2l}$ &  $107 \pm 5$  &** & $83 \pm 5$ & **& $75 \pm 5$&**\\
              & \textbf{108}  &** &\textbf{85} &** &\textbf{77} &** \\ \hline
     $M_{4l}$ &  $345 \pm 10$ &$235 \pm 23$ & $262 \pm 10$&$182 \pm 16$ &$212 \pm 20$& $143 \pm 17$ \\
              & \textbf{345}  &$\mathbf{230\pm 2}$ &\textbf{257} &$\mathbf{183\pm 2}$ &\textbf{200} & $\mathbf{148\pm 2}$\\ \hline
     $\overline{M}_{2l2l}$ &  $275 \pm 10$   &$141 \pm 10$ & $185\pm 15$&$109 \pm 10$ & $140\pm 20$& $86 \pm 11$\\
              & \textbf{282}  & $\mathbf{142\pm 2}$&\textbf{196} &$\mathbf{112\pm 2}$ &\textbf{156} &$\mathbf{91\pm 2}$\\ \hline
   $\overline{M}_{l3l}$ &  $290 \pm 10$  & $131 \pm 16$ &$163\pm 15$ &  $89 \pm 11$&$160\pm 20$ &  $69 \pm 10$\\
             & \textbf{ 288}  &$\mathbf{132\pm 5}$ &\textbf{180} &$\mathbf{89\pm 5}$ & \textbf{151}& $\mathbf{72\pm 2}$\\ \hline
   $\overline{M}_{l2l}$ &  $240 \pm 10$ & $111 \pm 10$& $139\pm 15$&$86 \pm 10$ & $135\pm 20$& $66 \pm 10$\\
             & \textbf{235} &$\mathbf{108\pm 3}$ &\textbf{147} &$\mathbf{89\pm 2}$ &\textbf{129} & $\mathbf{71\pm 2}$ \\ \hline
    $\overline{M}_{3l}$ &  $271 \pm 10$ &$164 \pm 11$ &$190\pm 10$ &$132 \pm 11$ & $150\pm 20$& $103 \pm 10$\\
             &  \textbf{266} &$\mathbf{168\pm 2}$ &\textbf{186} &$\mathbf{134\pm 2}$ &\textbf{148} & $\mathbf{107\pm 2}$\\ \hline
    $\overline{M}_{ll}$ &  $135 \pm 6$ &$72 \pm 5$ &$86\pm 8$ &$57 \pm 5$ & $90\pm 20$& $44 \pm 10$\\
             & \textbf{ 131} &$\mathbf{72\pm 1}$ & \textbf{86}&$\mathbf{58\pm 1}$ &\textbf{72} & $\mathbf{46\pm 1}$\\ \hline
   $|a_4|^{1/4}$ & **&$60 \pm 8$ &** &$50\pm 12$ & **&$41 \pm 10$ \\
             &** &$\mathbf{60\pm 1}$ &** &$\mathbf{42\pm 1}$ &** &$\mathbf{35\pm 1}$ \\ \hline
    \end{tabular}
    \end{center}
\end{table}

\subsection{Mass Extraction}

We now wish to  extract the masses  $m_A,~  m_{\widetilde{l}},~m_{{\widetilde\chi}^0_i},~ {\rm and}~ m_{{\widetilde\chi}^0_1}$
and their associated uncertainties from the MC data in Table \ref{tab:fits}.
First we search in mass space for solutions (within errors) to the
seven endpoint constraints using the formulae in the Appendix; in so doing we assume that the $M_{2l}$ edge has negligible error since if there is any modest rate for
${\widetilde\chi}^0_2 \to {\widetilde\chi}^0_1 l^\pm l^\mp$ via Higgs channels then surely there is a much greater rate via squarks\footnote{We may infer this, for example, from the presence of very jetty backgrounds.}, $e.g.$ $\tilde{q}_L \to q \widetilde{\chi}_{2}^0
\to \tilde{l}^\pm {l}^\mp q \to {l}^\pm  {l}^\mp  q \widetilde{\chi}_{1}^0$, so this edge will have been already measured very precisely (perhaps at the $0.1 \, \hbox{GeV}$ level\cite{massedge}).
We thus use the $M_{2l}$ formula to solve for $m_{{\widetilde\chi}^0_i}$ while scanning over  $m_A,~  m_{\widetilde{l}},~ {\rm and}~ m_{{\widetilde\chi}^0_1}$ for combinations that satisfy the other six endpoints within their assigned errors from Table \ref{tab:fits}.
As the reader can verify from the Appendix, the endpoint functions compose a highly nonlinear system of equations which, when each endpoint value has a small (percent level) uncertainty, generally gives rise to a discrete set of solutions (perhaps eight). In our case the uncertainty on each endpoint is somewhat larger, causing these discrete solutions to merge into a continuous range as
shown in the first row of Table \ref{tab:massfits}. However, many solutions with the same endpoints have very different (by say, tens of GeV) peaks\footnote{We know that theoretical peaks are always close to MC peaks (compared to error-bars) so this method will suffice; in a comparison between actual LHC data and theory we would use a more sophisticated estimate of peaks (at the level of the best available MC), or fit to whole distribution shapes.} --- though our peak resolutions are not always able to resolve these at SPS1a and Sample Point 2, we may narrow down the range of solutions somewhat at Sample Point 1 as seen in the second row of Table \ref{tab:massfits}.
If, in addition, the LSP mass is already known to fair precision ($e.g.$ $\pm 5 \, \hbox{GeV}$)
from inclusive measurements\cite{m1}  or other methods then   all masses at Sample Points 1 and 2 can be constrained well within 5\% of their nominal values. SPS1a does not fare as well (except for the Higgs mass, which is also within 5\%) but there are other techniques to constrain masses at this well-studied MSSM point which, when combined with ours, would certainly yield higher precision.

\begin{table}
 \caption{\small \emph{Extracted values of masses at MSSM Points 1,2 and SPS1a from matching theory to endpoints only ("EP"), endpoints and peaks ("EP+P"), or
  these in addition to constraining the LSP mass to within $5 \, \hbox{GeV}$ of its true value ("C.LSP").}}  \label{tab:massfits}
    \begin{center}
     \begin{tabular}{|c|c|c|c|c|} \hline
   Fit  & Parameter & MSSM 1  & MSSM 2 & SPS1a \\    \hline
     &$m_{{\widetilde\chi}^0_1}$& $136 \pm 54$ & $144 \pm 46$ & $135 \pm 55$ \\
      EP& $ m_{\widetilde{l}}$&$171 \pm 80$   & $185 \pm 51$  & $177 \pm 88$\\
        & $ m_{{\widetilde\chi}^0_i} $&$ 284 \pm 88$   & $230 \pm 47$ &$227 \pm 70$ \\
     & $m_A $&$ 627 \pm 123$   & $551 \pm 102$ & $473 \pm 121$ \\     \hline
        &$m_{{\widetilde\chi}^0_1}$&$137 \pm 53$   &  $144 \pm 46$ &  $135 \pm 55$ \\
      EP + P & $ m_{\widetilde{l}}$&$164 \pm 56$   & $185 \pm 51$  & $177 \pm 88$\\
        & $ m_{{\widetilde\chi}^0_i}$&$ 255 \pm  56$   & $230 \pm 47$& $222 \pm 65$\\
     & $m_A $&$ 623 \pm 119$    & $551 \pm 102$ & $473 \pm 121$ \\     \hline
       &$m_{{\widetilde\chi}^0_1}$&$\mathbf{128 \pm 5}$   & $\mathbf{121 \pm 5}$ &  $\mathbf{96 \pm 5}$\\
       C.LSP & $ m_{\widetilde{l}}$&$155 \pm 7$   & $156 \pm 5$  &$146 \pm 37$\\
         & $ m_{{\widetilde\chi}^0_i}$&$ 242 \pm  7$   & $208 \pm 3$ & $188 \pm 19$\\
     & $m_A $&$ 603 \pm 16$   & $497 \pm 11$ & $393 \pm 17$ \\     \hline
          \end{tabular}
    \end{center}
\end{table}

\section{Summary and Discussion}
\label{sec:sumdisc}

In the foregoing we have investigated seven invariant mass distributions
 ($M_{4l}$, $\overline{M}_{2l2l}$, $\overline{M}_{l3l}$, $\overline{M}_{l2l}$, $\overline{M}_{3l}$, $\overline{M}_{ll}$, and $a_4$)
 of the four-lepton endstate of the Higgs' decay (\ref{hdecay}), finding analytical expressions for endpoints of six of these ($M_{4l}$, $\overline{M}_{2l2l}$, $\overline{M}_{l3l}$, $\overline{M}_{l2l}$, $\overline{M}_{3l}$, $~ \mathrm{and }~\overline{M}_{ll}$) in the case of degenerate neutralinos ($i=j$) and sleptons.  Each invariant has three possible endpoint expressions depending on the precise ratios of sparticle masses; though we explicitly found these ratios for
  $M_{4l}$,  in practice one simply uses the maximum of the three endpoints as the correct one.
 Adding the well known dilepton invariant mass edge $M_{2l}$ to this list then gives seven invariant functions of  $m_A,~  m_{\tilde{l}},~m_{{\tilde\chi}^0_{i}},~ {\rm and}~ m_{{\tilde\chi}^0_1}$. We note this list of invariants is not unique: one could in principle select a totally different set (though still functions of (\ref{inv})) which might give sharper endpoints or peaks and improve upon the results we have presented here.

 One advantage of concentrating on the Higgs decay (\ref{hdecay}) is that one need not assume any specific choice of MSSM parameters: the  signal alone, $i.e.$ a boxlike wedgebox plot with a high ratio  $ \mathcal{R}_\pm$ of flavor balanced/unbalanced events identifies the responsible decays\footnote{Conversely, if the signal is absent then we can rule out a significant region of MSSM parameter space.}.
 One might then expect that any data sample of Higgs decays (\ref{hdecay}) with degenerate neutralinos and sleptons giving even modest determination of the seven endpoints would allow one  to overconstrain the four sparticle masses
  $m_A,~  m_{\tilde{l}},~m_{{\tilde\chi_i}^0},~ {\rm and}~ m_{{\widetilde\chi_1}^0}$.
 Were this determination more precise (at the GeV level) this would be so, at least up to a n-fold discrete ambiguity, but in realistic MC simulation we find these endpoints typically have a precision on the order of  $10 \, \hbox{GeV}$ or more --- in this case  masses can be constrained only within a large ($50-100\, \hbox{GeV}$) continuous range.
  Results improve with inclusion of peak constraints, but accuracies still linger at the 30\% level on average.
  If one of the masses ($e.g.$  $m_{{\widetilde\chi_1}^0}$) is already known with some precision then some or all of the other masses can be found with equal or better precision, depending on statistics and purity of sample ($\mathcal{R}_\pm$). In this way, Sample Points 1 and 2 lie in the center of disjoint regions in the $(\mu, M_2)$-plane where percent-level determination of all the masses is possible.
  More sophisticated analysis fitting to whole distribution shapes could in principle be employed for superior results.
 We could also couple this method with other techniques in the literature which do not employ invariant distributions\cite{massdet,massrel1}; such a 'hybrid'\cite{massrel2} analysis, particularly suitable to SPS1a for example,  would undoubtedly give much better mass determination.

 Our analysis could also be extended to colored sparticle decays where rates and therefore statistics are much higher.
 Here we have more avenues to explore: the gluino and squarks typically decay through multiple competing channels via more intermediate states, demanding a more intricate analysis; but we have jets in addition to leptons, so it may be possible to consider higher values of $N = n_j + n_\textit{l}$, giving many more invariants. The $(n_j,n_\textit{l})=(2,2)$ work of Ref.\cite{invmass3} is a promising step in this direction. In this connection, the concept  of the wedgebox plot, $i.e.$ a plot of di-electron versus di-muon invariant masses,
  could be extended to correlations between any two or three invariants which might prove as useful in isolating specific colored sparticle decays as
   we have found it to be invaluable in identifying the Higgs decays under study in the present work.

   Though MSSM parameters will ultimately be determined via a global fit to all available data, it is nevertheless important to have some model-independent idea of the rough values of these parameters as a starting point. We envision a general strategy of considering all possible $N = n_j + n_\textit{l}$ final states' kinematic invariant distributions ---
    endpoints, shapes, and correlations among these --- to not only single out specific decay chains but also pinpoint the masses in these. While up to now the community
    of SUSY phenomenologists had fruitfully executed this strategy up $N=4~(n_\textit{l}=2)$, in
    this work we have pushed the frontier to $N=4~(n_\textit{l}=4)$ with encouraging results.

\newpage

\newpage

\section*{Appendix}

We outline here the derivation of endpoint expressions (with the exception of $a_4$); to do this, we need to compute the four-momentum of each of the four leptons as functions of the eight spherical angles (defined in Fig.~\ref{decayfig}) in the frame of the decaying Higgs, and then apply (\ref{avinv}).

Let us start with the selectron, denoted $\tilde{l}_1$ in Fig.~\ref{decayfig}. In its rest frame, the LSP and electron  are produced back-to-back:
\begin{equation*}
p^\mu_1 =  \left(
                       \begin{array}{c}
                         E_1 \\
                         p_1 \sin \theta_1 \cos\phi_1 \\
                         p_1 \sin \theta_1 \sin\phi_1 \\
                         p_1 \cos \theta_1 \\
                       \end{array}
                     \right) ~~, ~~~~
p^\mu_{e^-} =  \left(
                       \begin{array}{c}
                         E_{e^-} \\
                         -p_1 \sin \theta_1 \cos\phi_1 \\
                         -p_1 \sin \theta_1 \sin\phi_1 \\
                         -p_1 \cos \theta_1 \\
                       \end{array}
                     \right)
\end{equation*}
where we have imposed momentum conservation ($\overrightarrow{p}_1+ \overrightarrow{p}_{e^-} = 0$) and we also have
\begin{equation*}
E_1 = \frac{m_s^2 + m_1^2}{2 m_s}~,~~  E_{e^-} = m_s - E_1~,~~ |p_1| = \frac{m_s^2 - m_1^2}{2 m_s}
\end{equation*}
 in an abbreviated notation:
  $m_i \equiv m_{{\widetilde\chi}^0_i} ~,~   m_1 \equiv m_{{\widetilde\chi}^0_1},$ and $m_s \equiv  m_{\tilde{l_1}}$.
Now we backwards Lorentz boost the electron along the z-direction by the selectron momentum (defined in the rest frame of the decaying
$\widetilde{\chi}^0_i$); the parameters of this boost are
\begin{equation*}
\gamma_1 = \frac{m_i^2 + m_s^2}{2 m_i m_s}~, ~~ \beta_1 \gamma_1 = \frac{m_i^2 - m_s^2}{2 m_i m_s}
\end{equation*}
and in this frame the positron has momentum
\begin{equation*}
p^\mu_{e^+} =  \left(
                       \begin{array}{c}
                         E_{e^+} \\
                        0  \\
                        0 \\
                         -E_{e^+} \\
                       \end{array}
                     \right)~, ~~   E_{e^+} = \frac{m_i^2 - m_s^2}{2 m_i}  
\end{equation*}
We now have the positron and electron momenta in the frame of the parent neutralino. To get these in the frame of the Higgs, we first rotate the coordinates so that the neutralino's momentum is along a new z-axis; this requires a rotation matrix of the form
\begin{equation*}
 \left(
                       \begin{array}{ccc}
   \cos\theta_i \cos\phi_i & -\sin\phi_i & \sin\theta_i \cos\phi_i \\
    \cos\theta_i \sin\phi_i & \cos\phi_i & \sin\theta_i \sin\phi_i \\
     -\sin\theta_i & 0 & \cos\theta_i \\
   \end{array}
                     \right)
\end{equation*}
And now we may backwards Lorentz boost by the neutralino momentum along z, using
\begin{equation*}
\gamma_i = \frac{m_i^2 - m_j^2 + m_A^2}{2 m_i m_A}~, ~~ \beta_i \gamma_i = 
\frac{\sqrt{(m_i^2 - m_j^2 + m_A^2)^2 - 4 m_i^2 m_A^2}}{2 m_i m_A}
\end{equation*}
The muon and antimuon momenta can be found analogously (being attentive to sign changes, of course) with the obvious substitutions of $i \to j$, $1 \to 2$, and $m_s \to m_{s'}$.

Endpoint expressions (\ref{m2l}) and (\ref{avinv}) are maximal when polar angles  $\cos{\theta_i}$,
$\cos{\theta_j}$, $\cos{\theta_1}$, and $\cos{\theta_2}$ are equal to $\pm 1$. 
For the  well-known dilepton mass edge, we of course have 
\begin{equation*}
    M_{2l}^{max} = m_i \sqrt{1-(m_s / m_i)^2}  \sqrt{1-(m_1 / m_s)^2}
\end{equation*}
Other endpoints have three maxima (labelled by polar angles  1 ('+') or -1 ('-') in brackets $[\pm \pm \pm \pm]$ as described in the text) which, for the case we consider where  $i=j$ and sleptons are degenerate,  are functions of $m_A$, $m_i$, $m_s$, and $m_1$.
The following definitions assist in writing each expression more compactly:
\begin{eqnarray*}
 \Delta  & \equiv &  \sqrt{m_A^4 - 4 m_A^2 m_i^2} \\
 J  & \equiv &  \sqrt{m_i^2 - m_s^2} \\
 F^\pm  & \equiv & \frac{1}{8}\left(m_A \pm  \sqrt{m_A^2 - 4 m_i^2} \right)^4 \\
 H^a_b   & \equiv & a ~m_i^2 - b ~m_s^2 \\
\end{eqnarray*}

{\small

\section*{}
\noindent\(
\mathbf{{M}_{4l}[++--]}= \frac{(m_i^2-m_1^2)
(m_A^2+\Delta)}{2 m_A m_i^2};\)
\\
\\
\noindent\(
\mathbf{{M}_{4l}[+---]}= \frac{1}{2 m_A m_i^2 
m_s^2}\left(m_1^2 m_i^2 (m_A^2+\Delta)-\Delta~
m_s^4+ m_A^2 (-2 m_i^2
m_s^2+m_s^4)\right);\)
\\
\\
\noindent\(
\mathbf{{M}_{4l}[----]}=  \frac{1}{2 m_A m_i^2
m_s^2} \left(m_1^2 m_i^2 (-m_A^2+\Delta)-\Delta~
m_s^4+ m_A^2 (2 m_i^2
m_s^2-m_s^4)\right);\)
\\
 \\
  \noindent\(
 \mathbf{\overline{M}_{2l2l}[-++-]}=
 \frac{1}{6^{1/4}}  \frac{(F^+)^{1/4}}{m_i^2}
{\huge \{ } 3 m_1^8 + m_i^6 H^3_{16} + 16  m_i^2 m_s^4 H^3_4 + 32 m_s^8
 + 4 m_1^6 H^1_4 + 6 m_1^4(m_i^2 H^3_8 + 8 m_s^4) + 4 m_1^2(m_i^4 H^1_{12} 
+  m_s^4  H^{24}_{16}) {\huge \}}^{1/4}
 ;\)
\\
\\
\noindent\(
 \mathbf{\overline{M}_{2l2l}[++--]}=\frac{1}{6^{1/4}}\frac{1}{m_i^2 m_s^2}
{\huge \{ }
3 m_1^8 m_i^8 F^+ - 4 m_1^6 m_i^8 m_s^2 (3 F^+ - J^2 [\Delta - 2 m_i^2 + m_A^2])
- 4 m_1^2 m_i^4 m_s^6 (3 m_A^4 m_i^4 + 32 m_i^8 + 24 m_i^4 m_s^4 - m_s^6 \Delta - 8 m_i^6 [\Delta + 6 m_s^2]
+ m_i^2[3 m_s^4 \Delta - 2 m_s^6] + m_A^2[m_s^6 - 16 m_i^6 - 3 m_i^2 m_s^4 + 3 m_i^4(\Delta + 2 m_s^2)])
+ 6 m_1^4 m_i^8 m_s^4 (3 F^+ + J^2 [10 m_i^2 - 6 m_s^2 - 2 \Delta - 2 m_A^2])
+ m_s^8 (3 m_A^2 J^8 [m_A^2 - \Delta] + 3 m_A^2 m_i^8 [m_A^2 + \Delta]- m_A^2 [32 m_i^6 J^4 + 52 m_i^4 m_s^4 J^2 + 12 m_i^2 m_s^8]  
+ 2 m_i^2 [32 m_i^{10} - 64 m_i^8 m_s^2 + 3 m_s^8 \Delta - 16 m_i^6 m_s^2(\Delta - 3 m_s^2) + 8 m_i^4(3 m_s^4 \Delta - 2 m_i^6)
+ m_i^2(3 m_s^8 - 14 m_s^6 \Delta)])
 {\huge \}}^{1/4}
; \)
\\
\\
\noindent\(
\mathbf{\overline{M}_{2l2l}[----]}= \frac{1}{6^{1/4}} \frac{1}{m_i^2 m_s^2}
{\huge \{ }
3 m_1^8 m_i^8 F^- + 4 m_1^6 m_i^8 m_s^2 ( J^2 [\Delta - 2 m_i^2 + m_A^2] - 3 F^-)
- 4 m_1^2 m_i^4 m_s^6 (3 m_A^4 m_i^4 + 32 m_i^8 + 24 m_i^4 m_s^4 + m_s^6 \Delta + 8 m_i^6 [\Delta - 6 m_s^2] 
- m_i^2[3 m_s^4 \Delta + 2 m_s^6] + m_A^2[m_s^6 - 16 m_i^6 - 3 m_i^2 m_s^4 - 3 m_i^4(\Delta - 2 m_s^2)])
- 6 m_1^4 m_i^8 m_s^4 (J^2 [10 m_i^2 - 6 m_s^2 - 2 \Delta - 2 m_A^2] - 3 F^-)
+ m_s^8 (3 m_A^2 J^8 [m_A^2 + \Delta] + 3 m_A^2 m_i^8 [m_A^2 - \Delta]- m_A^2 [32 m_i^6 J^4 + 52 m_i^4 m_s^4 J^2 + 12 m_i^2 m_s^8]  
+ 2 m_i^2 [32 m_i^{10} - 64 m_i^8 m_s^2 - 3 m_s^8 \Delta + 16 m_i^6 m_s^2(\Delta + 3 m_s^2) - 8 m_i^4(3 m_s^4 \Delta + 2 m_i^6)
+ m_i^2(3 m_s^8 + 14 m_s^6 \Delta)])
{\huge \}}^{1/4}
;\)
\\
\\
\noindent\(
\mathbf{\overline{M}_{l3l}[+--+]}= \frac{1}{2^{1/4}} \frac{\sqrt{m_s^2 - m_1^2}}{m_i m_s}
{\huge \{ }
F^+(m_1^2 - m_s^2)^2 + J^4 (F^+ + 2 m_i^4)
{\huge \}}^{1/4}
;\)
\\
\\
 \noindent\(
   \mathbf{\overline{M}_{l3l}[--+-]}=  \frac{1}{2^{1/4}} \frac{J}{m_i^2 m_s} { \huge \{} m_s^4[(m_s^2 - m_1^2)^2 + J^4] F^+ + 2 m_i^8 (m_1^2 - m_s^2)^2 { \huge \}}^{1/4}
;\)
\\
  \\
  \noindent\(
 \mathbf{\overline{M}_{l3l}[++--]}=
 \frac{1}{2^{1/4}} \frac{1}{m_i^2 m_s^2}
{\huge \{ }
F^+(m_1^8 m_i^8 - 4 m_1^6 m_i^8 m_s^2) + 2 m_1^4 m_i^8 m_s^4(3F^+ - 2 J^4)
- 4 m_i^2 m_i^8 m_s^6(F^+ - 2 J^4) + m_s^8(J^8[\Delta^2 - m_A^2 \Delta] + m_i^8[\Delta^2 + m_A^2 \Delta])
{\huge \}}^{1/4}
;\)
\\
\\
\noindent\(
\mathbf{\overline{M}_{l2l}[-++-]}= \frac{1}{\sqrt{2} ~ 3^{1/4}} \frac{1}{m_i^2} (F^+)^{1/4} 
{\huge \{ } 3 m_1^8 - m_i^6 H^{14}_3 -2 m_s^4 m_i^2 H^{16}_{15} +16 m_s^8 +2 m_1^6 H^1_7 + 6 m_1^4( m_i^2 H^4_3 + 5 m_s^4) +12 m_1 ^2 (m_i^4 H^1_9 + 2 m_s^4 H^9_8) 
{\huge \}}^{1/4}
;\)
\\
\\
\noindent\(
\mathbf{\overline{M}_{l2l}[++--]}= \frac{1}{\sqrt{2}~ 3^{1/4}} \frac{1}{m_i^2 m_i^2} { \huge \{} 3 m_1^8 m_i^8 F^+ - 2 m_1^6 m_i^8 m_s^2 [ (2 m_i^2 - \Delta - m_A^2)J^2 + 6 F^+] + 6 m_1^4 m_i^8 m_s^4 [ 3 F^+ + (2 H^2_1 - \Delta - m_A^2 )J^2 ] - 2 m_1^2 m_i^4 m_s^6 [ 6 m_A^4 m_i^4 + 32 m_i^8 - 18 m_i^4 m_s^2 H^2_1 - \Delta m_s^6 -14 m_i^6 \Delta +3 m_i^2 m_s ^4 \Delta - 2 m_s^6 m_i^2 + m_A^2( -2 m_i^4 H^{14}_3 -3 m_s^4 H^1_2 + 6 m_i^4 \Delta) ] + m_s^8[ 3 m_A^4 (J^8 + m_i^8) + m_A^2( -3 J^8 \Delta +3 m_i^8 \Delta - 28 m_i^6 J^4 - 50 m_i^4 m_s^4 J^2 - 12 m_i^2 m_s^8) + 32 m_i^{10} H^1_2 + 6 m_i^2 m_s^8 \Delta + m_i^8( -28 m_s^2 \Delta + 60 m_s^4) + 2 m_i^4 m_s^4 \Delta H^{21}_{13} -2 m_i^4 m_s^6 H^{14}_3 ] { \huge \}}^{1/4}
;\)
\\
\\
\noindent\(
\mathbf{\overline{M}_{l2l}[----]}= \frac{1}{\sqrt{2}~ 3^{1/4}} \frac{1}{m_i^2 m_i^2} { \huge \{} 3 m_1^8 m_i^8 F^- + 2 m_1^6 m_i^8 m_s^2 [ (-2 m_i^2 - \Delta + m_A^2)J^2 - 6 F^-] + 6 m_1^4 m_i^8 m_s^4 [-3 F^- + (-2 H^2_1 - \Delta + m_A^2 )J^2 ] - 2 m_1^2 m_i^4 m_s^6 [ 6 m_A^4 m_i^4 + 32 m_i^8 - 18 m_i^4 m_s^2 H^2_1 + \Delta m_s^6 -14 m_i^6 \Delta - 3 m_i^2 m_s ^4 \Delta - 2 m_s^6 m_i^2 + m_A^2( -2 m_i^4 H^{14}_3 -3 m_s^4 H^1_2 - 6 m_i^4 \Delta) ] + m_s^8[ 3 m_A^4 (J^8 + m_i^8) + m_A^2 ( 3 J^8 \Delta - 3 m_i^8 \Delta - 28 m_i^6 J^4 - 50 m_i^4 m_s^4 J^2 - 12 m_i^2 m_s^8) + 32 m_i^{10} H^1_2 - 6 m_i^2 m_s^8 \Delta + m_i^8( 28 m_s^2 \Delta + 60 m_s^4) + 2 m_i^4 m_s^4 \Delta H^{21}_{13} -2 m_i^4 m_s^6 H^{14}_3 ] { \huge \}}^{1/4}
;\)
\\
 \\
\noindent\(
 \mathbf{\overline{M}_{3l}[-++-]}=  \frac{1}{\sqrt{2}} \frac{J}{m_i^2} (F^+)^\frac{1}{4} {\huge \{} (m_s^2 - m_1^2)^2 + J^4 {\huge \}}^{1/4}
;\)
\\
 \\
\noindent\(
\mathbf{\overline{M}_{3l}[++--]}= \frac{1}{\sqrt{2}} \frac{1}{m_i^2 m_s^2} {\huge \{} m_1^8 m_i^8 F^+ + 2 m_1^6 m_i^8 m_s^2 [J^2 (2 m_i^2 - \Delta - m_A^2) - 2 F^+ ] + 2 m_1^4 m_i^8 m_s^4 [3 F^+ + J^2 (3 m_A^2 +2 \Delta - 2 H^2_{-1})] - 2 m_1^2 m_i^4 m_s^6[2 m_i^4(m_A^4 + m_s^4) + m_s^6 \Delta - 2 m_i^6(\Delta - 2 m_s^2) + m_i^2(-3 m_s^4 \Delta + 2 m_s^6) + m_A^2(m_s^4 H^3_1 - 2 m_i^4 H^2_3 + 2 m_i^4 \Delta)] + m_s^8[ (\Delta^2 - m_A^2\Delta)J^8 + m_i^8(m_A^4 + m_A^2 \Delta - 8 m_A^2 m_s^2) + 2 m_A^2 m_i^4 m_s^4 H^3_1 - 2 m_i^6 m_s^2 \Delta H^2_3 + 4 m_s^4 m_i^6 H^1_1 - 2 m_i^2 m_s^6 \Delta H^3_1 + 2 m_i^4 m_s^8] {\huge \} }^{1/4};\)
\\
\\
\noindent\(
 \mathbf{\overline{M}_{3l}[----]}= \frac{1}{\sqrt{2}} \frac{1}{m_i^2 m_s^2} {\huge \{} m_1^8 m_i^8 F^- + 2 m_1^6 m_i^8 m_s^2 [J^2 (2 m_i^2 + \Delta - m_A^2) - 2 F^- ] + 2 m_1^4 m_i^8 m_s^4 [3 F^- + J^2 (3 m_A^2 - 3 \Delta - 2 H^2_{-1})] - 2 m_1^2 m_i^4 m_s^6[2 m_i^4(-m_A^4 + m_s^4) + m_s^6 \Delta - 2 m_i^6(\Delta - 2 m_s^2) + m_i^2(3 m_s^4 \Delta + 2 m_s^6) + m_A^2(-m_s^4 H^3_1 + 2 m_i^4 H^2_3 + 2 m_i^4 \Delta)] + m_s^8[ (\Delta^2 + m_A^2\Delta)J^8 + m_i^8(m_A^4 - m_A^2 \Delta - 8 m_A^2 m_s^2) + 2 m_A^2 m_i^4 m_s^4 H^3_1 - 2 m_i^6 m_s^2 \Delta H^2_1 + 4 m_s^4 m_i^6 H^1_1 + 2 m_i^2 m_s^6 \Delta H^1_1 + 2 m_i^4 m_s^8] {\huge \} }^{1/4}
;\)
\\
\\
\noindent\(
\mathbf{\overline{M}_{ll}[-++-]}= \frac{1}{2~ 3^{1/4}} \frac{1}{m_i^2} (F^+)^{1/4}\sqrt{(m_s^2 - m_1^2)^2 + J^4}
;\)
\\
 \\
\noindent\(
\mathbf{\overline{M}_{ll}[----]}= \frac{1}{2 ~3^{1/4}} \frac{1}{m_i^2 m_s^2} { \huge \{ } m_1^2 m_i^8 (m_1^6 - 4 m_1^4 m_s^2 +6 m_1^2 m_s^4 - 4 m_s^6) F^- - 4(2 m_s^2 - m_1^2) m_1^2 m_i^8 m_s^4 J^4 + m_s^8 [ \Delta (J^8 - m_i^8) (\Delta + m_A^2) + 2 m_i^2 (J^4 + m_i^4)( m_i^2 J^4 + m_i^6 - m_s^2 \Delta H^2_1) ] {\huge \}}^{1/4}
;\)
\\
 \\
\noindent\(
\mathbf{\overline{M}_{ll}[++--]}= \frac{1}{2~ 3^{1/4}} \frac{1}{m_i^2 m_s^2} { \huge \{ } m_1^2 m_i^8 (m_1^6 - 4 m_1^4 m_s^2 +6 m_1^2 m_s^4 - 4 m_s^6) F^+ - 4(2 m_s^2 - m_1^2) m_1^2 m_i^8 m_s^4 J^4 + m_s^8 [ \Delta^2 (J^8 + m_i^8) -\Delta m_A^2(J^8 - m_i^8) + 2 m_i^2 (J^4 + m_i^4)(J^4 m_i^2 + m_i^6 - m_s^2 \Delta H^2_1) ] {\huge \}}^{1/4}
; \)
}

\bigskip

In the event that decays proceed through off-shell sleptons, the endpoint expressions simplify dramatically, $e.g.$ for $M_{4l}$, $\overline{M}_{2l2l}$,$ \overline{M}_{l2l}$, $\overline{M}_{3l}$, $~ \mathrm{and }~\overline{M}_{ll}$, either $[++--]$ or $[--++]$ (alternatively $[----]$ or $[++++]$, respectively) are maximal. 
For $i=j$, these are:
\begin{eqnarray*}
  M_{4l}[++--] &=& m_A \left( 1 - \frac{m_1}{m_i} \right) \\ 
  M_{4l}[--++] &=& \frac{m_A^2 + \Delta}{2 m_A} \left( 1 - \left(\frac{m_1}{m_i}\right)^2 \right) \\ 
\overline{M}_{2l2l}[++--] &=&  \left( 1 - \frac{m_1}{m_i} \right)
                 \left(\frac{3 m_A^4 - 16 m_A^2 m_i^2 + 32 m_i^4}{3}\right)^{1/4} \\
\overline{M}_{2l2l}[--++] &=&  M_{4l}[--++] \\
\overline{M}_{l2l}[++--] &=&  \left( 1 - \frac{m_1}{m_i} \right)
                 \left(\frac{3 m_A^4 - 14 m_A^2 m_i^2 + 16 m_i^4}{6}\right)^{1/4} \\
\overline{M}_{l2l}[--++] &=&  2^{-1/4}M_{4l}[--++] \\
\overline{M}_{3l}[++--] &=&  \left( 1 - \frac{m_1}{m_i} \right)
                 \left(\frac{3 m_A^4 - 2 m_A^2 m_i^2}{2}\right)^{1/4} \\
\overline{M}_{3l}[--++] &=&  2^{-1/4}M_{4l}[--++] \\
\overline{M}_{ll}[++--] &=&  \frac{1}{24^{1/4}} \left( 1 - \frac{m_1}{m_i} \right)
                             \sqrt{m_A^2 - 2 m_i^2} \\
\overline{M}_{ll}[--++] &=&   \frac{1}{2}  \frac{1}{6^{1/4}}
     \left( 1 - \left(\frac{m_1}{m_i}\right)^2 \right)
      \sqrt{m_A^2 - 2 m_i^2 + \Delta} \\
\end{eqnarray*}
It should be easy to see the difference between this pattern of endpoints and those for the on-shell case.

\end{document}